\documentstyle[proc]{rspublic}

\begin{document}
\input{epsf}
\title[Testing Top]{Test of Analysis Method for
Top-Antitop Production and
Decay Events}
\author[R.H.Dalitz and G.R.Goldstein]{R. H. Dalitz$^1$ and Gary R.
Goldstein$^2$}
\affiliation{1. Department of Physics (Theoretical Physics)
University of Oxford \\
1 Keble Road, Oxford OX1 3NP  UK \\
and \\
2. Department of Physics
Tufts University
Medford, MA 02155  USA}
\maketitle
\label{firstpage}
\begin{abstract}
We have carried out Monte Carlo calculations on two sets of randomly 
generated QCD events due to $p\bar{p}\,\rightarrow\, t\,
\bar{t}$ with top mass $m_t = 170$ GeV, one set leading to
$e^+e^-$ or $e^{\pm}\mu^{\mp}$ or $\mu^+\mu^-$ 
2jets (dilepton) and the other leading to $e^{\pm}$ or
$\mu^{\pm}$ 4jets (unilepton) configurations, in order to test the 
likelihood methods we have proposed for determining the top mass by 
analyses of these two sets of configurations. For
the set of unilepton events, our method gives a very efficient and quite
sharp measure of the top mass lying several GeV below the input mass. For
the dilepton set, our method gives a much broader and markedly asymmetric
distribution for the top mass estimates, 75\% of them lying below 170 GeV,
but the dilepton data will have much lower background than unilepton data.
We then illustrate these methods by applying them to the data available from 
CDF in 1995  and discuss the results obtained in relation to the results 
for the sets of Monte Carlo events. The dilepton events yield 
masses spread widely, over 140 to 180 GeV, generally lower 
than the unilepton events, which cluster around $175\pm 8$ GeV.
In an appendix, we discuss the nature of the
additional ``slow'' $\mu^{+}$ observed in one CDF dilepton event,
concluding that it is most probably a ``tertiary lepton'' resulting from
the decay sequence $b\rightarrow c +\mathrm{hadrons}$, followed by
$c\rightarrow s \mu^{+} \nu$.
\end{abstract}


\section{INTRODUCTION}

    A major step forward in research on the top quark has recently been
achieved by the CDF Collaboration (Abe {\em et al.} 1994a, 1995a,b) and
the D0 Collaboration (Abachi {\em et al.} 1995). Both groups present
evidence for the discovery
in events interpreted as being due to the following production and decay 
sequences, observed in experiments with the Tevatron at the Fermi National 
Accelerator Laboratory at Batavia (Illinois):
\begin{equation} \bar{p} + p \rightarrow \bar{t} + t + {\rm other \: hadrons},
\label{pbarp}
\end{equation}
which, according to our theoretical picture of the proton, is initiated,
at the subnuclear level, mainly by the quark-antiquark annihilation
reaction,
\begin{equation} \bar{q} + q \rightarrow \bar{t} + t,
\label{qbarq}
\end{equation}
where q denotes a valence quark, u or d, in the proton. This is 
followed by the rapid decay processes for the top and antitop quarks
(lifetime width $\simeq$1.4 GeV),
\begin{equation}
   t \rightarrow W^{+} b \ \ \ {\rm and}\ \ \ \bar{t} \rightarrow W^{-} \bar{b} 
\label{twb} 
\end{equation}
with 
\begin{equation}
(a) \ \ W^+ \rightarrow l^+\nu_l \ \ \ {\rm or}\ \ \ (b) \ \ W^+ \rightarrow
 u\bar{d} \:{\rm or}\: c\bar{s}
\label{wplus} 
\end{equation}
and
\begin{equation}
(a) \ \ W^- \rightarrow l^-\bar{\nu}_l \ \ \ {\rm or}\ \ \ (b) \ \  W^-
\rightarrow d\bar{u} \: {\rm or} \: s\bar{c}. 
\label{wminus} 
\end{equation} 

For the $W^+$ (or $W^-$) decays, the hadronic modes(\ref{wplus}b) (or 
(\ref{wminus}b)) have a net rate of about 9 times the rate for each 
leptonic mode. It is therefore not surprizing that the bulk of the present 
evidence on the top quark arises from the unilepton events,
\begin{equation}
 t \bar{t}\rightarrow b \bar{b}\, (\bar{q} q)\, l \nu_l
\label{unievent}
\end{equation}
rather than from the dilepton events
\begin{equation}
 t \bar{t}\rightarrow b \bar{b}\, l \bar{l}\, \nu_l \bar{\nu}_l,
\label{dievent}
\end{equation}
where $l \bar{l} = (e^+e^-),\, (e^{\pm}\mu^{\mp}),\, {\rm and}\, (\mu^+ 
\mu^-)$.
Events of the type 
\begin{equation}
 t \bar{t}\rightarrow b \bar{b} (\bar{q} q) (\bar{q} q)
\label{hadevent}
\end{equation}
are dominant, of course, but they lead to six final jets, a complicated
final state to unscramble, which we do not discuss in this paper: however,
see Benlloch, Wainer and Giele (1993) for an optimistic assessment of the
latter. A preliminary result has been given by Abe, {\textit {et al.}}
(1997).

Both groups have put forward estimates of $m_t$, the top quark mass, on 
the basis of their own data. The official best estimate (Particle Data  
Group 1996) is
\begin{equation}
  m_t=180 \pm 12 {\rm GeV},
\label{pdgmass}
\end{equation}
this value being dominated by the CDF result. Updated estimates for $m_t$
were given by both groups at the recent 18th International
Electron-Photon Conference held in Hamburg (Giromini 1997), as follows:
\begin{equation}
(a)\, {\mathrm{CDF}} \ m_t=176.8 \pm 6.5 \ {\mathrm{GeV}}, \ (b)\, 
{\mathrm{D0}} \ m_t=173.3 \pm 8.4 \ \mathrm{GeV}.
\label{newmt}
\end{equation}

Over several years, we (Dalitz \& Goldstein 1992a,b,1994   
and Goldstein, Sliwa \& Dalitz 1993)
and K.~Kondo (Kondo 1988,1991, and Kondo, Chikamatsu \& Kim 1993)
independently, have
developed a method for determining whether events 
of the type reported by CDF and D0 Collaborations are consistent
with the hypothesis that they are examples of top-antitop pair creation
and their decay through the steps (\ref{twb}), (\ref{wplus}) and
(\ref{wminus}) given above, leading to final states (\ref{unievent}) and
(\ref{dievent}).

The procedure is to take the measured configuration of momenta for the
final leptons and jets in a single event $i$ and to evaluate the
probability
\linebreak $P_i(m)=P({\mathrm{configuration \ event}}\ i|m)$ that these
production and decay
processes could produce the observed configuration if the top quark mass
were $m$. This evaluation must take into account each step in the
processes (\ref{pbarp}) through (\ref{wminus}).

(a) The initial partons $q$ and $\bar{q}$ have momenta $x\textbf{P}$ and
$-\bar{x}\textbf{P}$, where $\textbf{P}$ denotes the incident proton
momentum in the proton-antiproton rest-frame. The values $(x,\bar{x})$
for
$t\bar{t}$ production at the Tevatron are so large (Dalitz \& Goldstein
1992b,1994) that $q$ and $\bar{q}$ are dominantly valence quarks. The
proton structure function $F(x)$ for valence quarks is well-known in the
regime which is important for the process (\ref{qbarq}); the antiproton
structure function is $\bar{F}(\bar{x})=F(x)$.

(b) The complete ($t\bar{t}$) sustem and its final states are to be viewed
in the ($t\bar{t}$) rest-frame achieved by a suitable boost along
$\textbf{P}$. The amplitude for process (\ref{qbarq}) will depend on the
angle $\theta$ between the t-momentum in this rest-frame and the direction
$\textbf{P}$ and on the total rest energy $m(t\bar{t})$ of the
$t\bar{t}$-system produced. With Quantum Chromodynamics (QCD), this
amplitude is usually taken to be that produced by the simplest QCD graph,
here that for $q\bar{q}\rightarrow g \rightarrow t\bar{t}$, and we have
adopted this in our work. We note that this mechanism implies a strong
spin-spin correlation between $t$ and $\bar{t}$. For subsequent decays,
the angles for various decay products will be defined relative to the
plane defined by the $t\bar{t}$ production plane in the lab frame.

(c) The decay $t\rightarrow bW^+$ is specified in the t rest-frame, by
angles $\theta_W^+$ and $\phi_W^+$. $\theta_W^+$ is the angle between
the momentum \textbf{p}$_{W^+}$ in this frame and the boost direction
\textbf{z}$_t$ from the
$t\bar{t}$ rest-frame to this t rest-frame. $\phi_W^+$ is the azimuth
angle of \textbf{p}$_{W^+}$ relative to the production plane and the
\textbf{z}$_t$ axis. Similarly, to
specify $\bar{t}\rightarrow \bar{b}W^{-}$, there are corresponding angles
 $\theta_W^{-}$ and $\phi_W^{-}$. With \textbf{z}$_{\bar{t}}$ being
the boost direction from the $t\bar{t}$ rest-frame to the $\bar{t}$
rest-frame, $\theta_W^{-}$ is the angle between the momentum
\textbf{p}$_{W^{-}}$ and \textbf{z}$_{\bar{t}}$, while $\phi_W^{-}$
is the azimuth angle of \textbf{p}$_{W^{-}}$ relative to the production
plane and the \textbf{z}$_{\bar{t}}$ axis.
 
(d) The decay $W^+\rightarrow l^+\nu_l$ is specified in the $W^+$ 
rest-frame by angles $\theta_l^+$ and $\phi_l^+$. $\theta_l^+$ is the angle
between the lepton momentum \textbf{p}$_{l^+}$ in this frame and the boost
direction \textbf{z}$_{W^+}$ from the t rest-frame
to the $W^+$ rest-frame. $\phi_l^+$ is the azimuth angle of
\textbf{p}$_{l^+}$ relative to the plane formed by \textbf{z}$_t$ and
\textbf{p}$_{W^+}$ in the t rest-frame. The non-leptonic decays
$W^+\rightarrow u\bar{d}\:
{\mathrm{or}}\: c\bar{s}$ can be specified by corresponding angles
($\theta_l^+$, 
$\phi_l^+$), where the positive lepton label is replaced by the u
or c quark. Since we have no ready means to
distinguish a $u$ from a $\bar{d}$ jet, nor $c$ from $\bar{s}$ jet, for  
pairs of jets, we have to add the rates for the angles $\theta_q^+$
and $(\pi-\theta_q^+)$ for $q\ =\ u$ or $c$. Similarly, to specify
$W^-\rightarrow l^-\bar{\nu}_l$ in the $W^-$ rest-frame, there are
corresponding angles $\theta_l^-$ and $\phi_l^-$ whose definitions involve
replacing $t$ with $\bar{t}$ and $W^+$ with $W^{-}$ in all of the above.
Furthermore there are corresponding remarks that are appropriate for the
decays $W^-\rightarrow d\bar{u} \: {\mathrm{or}} \: s\bar{c}$.

Thus, each $t\bar{t}$ final configuration requires the specification of
eleven numbers
\begin{equation}
\left\{ \begin{array}{cccc}
  x &  & (\theta_W^+,\phi_W^+) & \left\{ \begin{array}{ccccc}
(\theta_l^+,\phi_l^+) & {\mathrm{or}} & (\theta_u^+,\phi_u^+) &
{\mathrm{or}} & (\theta_c^+,\phi_c^+) \end{array} \right\} \\
    & \theta & & \\
\bar{x} & & (\theta_W^-,\phi_W^-) & \left\{ \begin{array}{ccccc}
(\theta_l^-,\phi_l^-) & {\mathrm{or}} & (\theta_u^-,\phi_u^-) &
{\mathrm{or}} & (\theta_c^-,\phi_c^-) \end{array} \right\} 
\end{array}  \right\} 
\label{angles}
\end{equation}
if we sum over the spins of the initial $q$ and $\bar{q}$, the spins of
the $t$ and $\bar{t}$, 
the spins of the final $b$ and $\bar{b}$, and the spins of the final
leptons $l^+$ and
$l^-$ (or the spins of the quarks from non-leptonic decays). We note
that, with the Standard Model, all the transitions
occurring subsequent to the creation of the $t\bar{t}$ system are
completely prescribed in form and magnitude. These transitions have been
fully described in an earlier publication (Dalitz \& Goldstein 1992b). We
emphasize that our calculations are carried out while retaining the spin
and tensor polarizations of the $W^{\pm}$ mesons, and repeat that we have
always averaged over $t$ and $\bar{t}$ polarizations. 

The final step involves the use of Bayes' Theorem, which gives the
probability distribution of mass $m$, given data on a set of events
$\{i\}$. This theorem states that for an event $i$
\begin{equation}
P(m|{\mathrm{data\ on\ event}}\: i)=P({\mathrm{observed \ event}} \: 
i|m)\cdot \Phi(m), 
\end{equation}
where $\Phi(m)$ represents the $\textit{a priori}$ probability that the
top mass is $m$. For a set $\{i\}$ of N events, this implies that
\begin{equation}
P(m|{\mathrm{data \ set}}\: \{i\}) =\prod_{i=1}^{N}
P({\mathrm{event}}\: i|m)\cdot \Phi(m).
\end{equation}
In the following we use the notation $P_i(m)$ for the
distribution $P({\mathrm{event}}\: i|m)$
for each individual event and we parametrise these distributions in a simple
way, using two parameters, $m_{pk}(i)$ and $LIP(i)$, defined below in Sec.2.
We then determine how these and other possible parameters behave over a set
of events.

Our purpose here has been to determine what behaviour we should expect
for $P(m|{\mathrm{data \ set}}\ \{i\})$ when our analysis method is
applied to
a large ($\approx
100$) batch of $t\bar{t}$ production and decay events. This knowledge
will allow us to
discriminate between events which are due to production and decay of
$t\bar{t}$ systems from other events whose final states have the same
leptons and quark jets, but which do not involve intermediate $t$ and
$\bar{t}$ particles. Lacking guaranteed $t\bar{t}$ production and decay
events on which to test our method, we have had to generate a set of such
events by computer simulation and to use them to test the method.

In Sec.2, we describe the procedure we followed to generate sufficiently
large random samples of these events, using only the very simplest
$q\bar{q}\rightarrow t\bar{t}$ QCD graph, and the results found from our
analysis of these events. The remarkable difference we find there between
the outcome for dilepton events and the outcome for unilepton events
(see Fig.3),
appears to be in good qualitative accord with current experimental data
(see Figs.6 and 9). In Sec.3, we illustrate our method by examining 3
dilepton events which have been reported in the literature (Abe \textit{et
al.} 1994b), commenting further on the additional 11 events which have
become
available since the 1997 Electron-Photon conference at Hamburg in July
(Gironimi 1997). In
Sec.4, we illustrate our method by examining seven unilepton events
reported by CDF (Abe \textit{et al.} 1994b), with conclusions in good
general
accord with those of CDF, except for one or two doubtful events, which
might well not be due to $t\bar{t}$ production and decay. In an Appendix
we discuss an anomalous event in which a ``secondary lepton'' has low
$p_T$ for only one of the four jets, but has the wrong charge sign for
this interpretation, and make out a quantitative case for regarding this
lepton to be a ``tertiary lepton''.

\section{Monte Carlo Tests of our Analysis Procedures}

With the likelihood methods we have proposed for the analysis of 
${\bar t} t$ production and decay events to determine the physical
top mass $m_t$, it is essential for us to understand the quantitative
significance of the values determined for the probability function
$P_i(m)$ from an experimental event labelled $i$.  If a set of $N$
events are all top-antitop events, the peak mass value $m_{pk}$ for the
product
\begin{equation}
P(m) = \prod_{i=1}^N P_i(m)
\end{equation}
will be the Bayesian estimate for the top mass $m_t$.  However,
we require more than this.  The peak probability values $P_i(m_{pk}(i))$
for each event should have acceptable magnitudes and they must have a
reasonable
distribution over a set of $N$ events. What these should be could best be
established by making 
an analysis over some large batch of guaranteed events of this type.
For this purpose, the only procedure available to us is to develop
random sets of computer-generated events using a Monte Carlo (MC)
simulation based on tree-level QCD Feynman graphs for top-antitop
production, followed by their decay sequences (1.3) (1.4) and (1.5) to
reach the final states ``$l^\pm$ 4 jets'' and ``$l^\pm l^{\prime\mp}$ 2
jets'', in which the former has also one neutrino and the latter, where
$l^{\prime}$ may or may not be the same as $l$, one  neutrino and one
antineutrino, unmeasured, and to carry out our analysis procedures on
them,
after making allowances for the proper development of the jets and the 
energy resolutions appropriate to the measurements of all the final
particles.  One may wonder whether such a simple ansatz as the 
single-gluon graph for $\bar q q \rightarrow gluon \rightarrow \bar t t$
should be adequate for the description of top-antitop pair production
in a proton-antiproton interaction at 1.8 TeV.  Well, it is at least
the simplest possible process existing within QCD, and a number of other
proton-antiproton processes, not involving top quarks, have given data
in quite a good agreement on the basis of cross sections calculated
using QCD matrix-elements having the same degree of simplicity

\subsection{ The analysis procedure for unilepton events}

We began by generating 100,000 configurations for a mass $m_0 = 170$
Gev assumed for the top quark, and divided them successively into five
batches of 20,000.  As remarked above, each configuration, say $i$, with
$i$ running from 1 to 100,000, requires the specification of the eleven
possible variables (\ref{angles}), each of which varies over a finite
range.
If the lepton has charge +1, then ($\theta_l^+,\phi_l^+$) gives its
direction
of emission in the $W^+$ rest frame, determined as specified in Sec.1.
The corresponding $W^-$ decays to ($\bar{u},d$) or ($\bar{c},s$), with 
($\theta_u^-,\phi_u^-$) or ($\theta_c^-,\phi_c^-$) now representing the
direction of the $\bar{u}$ or
$\bar{c}$ quark in the $W^-$ rest frame. 
If the lepton has negative charge, then 
($\theta_l^{-} , \phi_l^{-}$) specifies its direction of
emission in the
$W^-$ rest frame, while ($\theta_u^-,\phi_u^-$) or ($\theta_c^-,\phi_c^-$)  
gives the direction of the
u or c quark in the $W^+$ rest frame.  The finite 11-dimensional
space of these variables was divided into a finite number of cells (Barger
and Phillips 1987).  Each event is assigned a weight $w_i$, which is equal
to its
fractional contribution to the theoretical total cross section.  The 
aim is to make the cells about equal in net weight, summed over all the
events in the cell, by changing the cell parameters as necessary.  There
is a convenient iterative procedure available, based on a program by
Ohnemus (see Barger, Han, Ohnemus \& Zeppenfeld 1990).
 
The first batch of 20,000 events led to a net weight in each cell, and
a new set of partitions of the eleven variables were chosen, following
Ohnemus, leading
to a second set of cells, each with about the same weight.  The next
20,000 events (batch 2) were then chosen and located in the new
cells, leading to differing weights in these cells, so that a third
set of partitions of the eleven variables had to be chosen, to even 
up the cell weights again.  This procedure was iterated five times,
for each new batch of randomly generated events.  The distribution
of the fifth batch of 20,000 events in the 11-dimensional space is 
then expected to be much closer to the physical reality corresponding
to the simple tree-level model, than that of the first 20,000 events.
For each batch of configurations, and the events they give rise to,
their number was reduced by requiring each event to satisfy experimental
cuts that approximated those used by CDF for unilepton events.
This reduces the size of each batch substantially.  The cuts chosen
required a minimum of 10 and 20 GeV/c transverse momentum for hadronic
jets and leptons, respectively, a minimum pseudorapidity separation
from one outgoing particle to another of 2.5 units, and a minimum
separation between jets of 0.7 units in the pseudorapidity-azimuthal
angle plane.  Now the improvement achieved in each iteration can be
tested, for example by comparing the total cross section calculated after
each iteration with the directly calculated total cross section
for this simple model.

We then chose arbitrarily all those events from batch 5 which survived
the cuts applied experimentally to unilepton events and which bore a 
number between 80,000 and 83,000.  These events were 1292 in number,
each with its own $w_i$.  We 
\begin{figure}
\begin{center}\leavevmode\epsffile{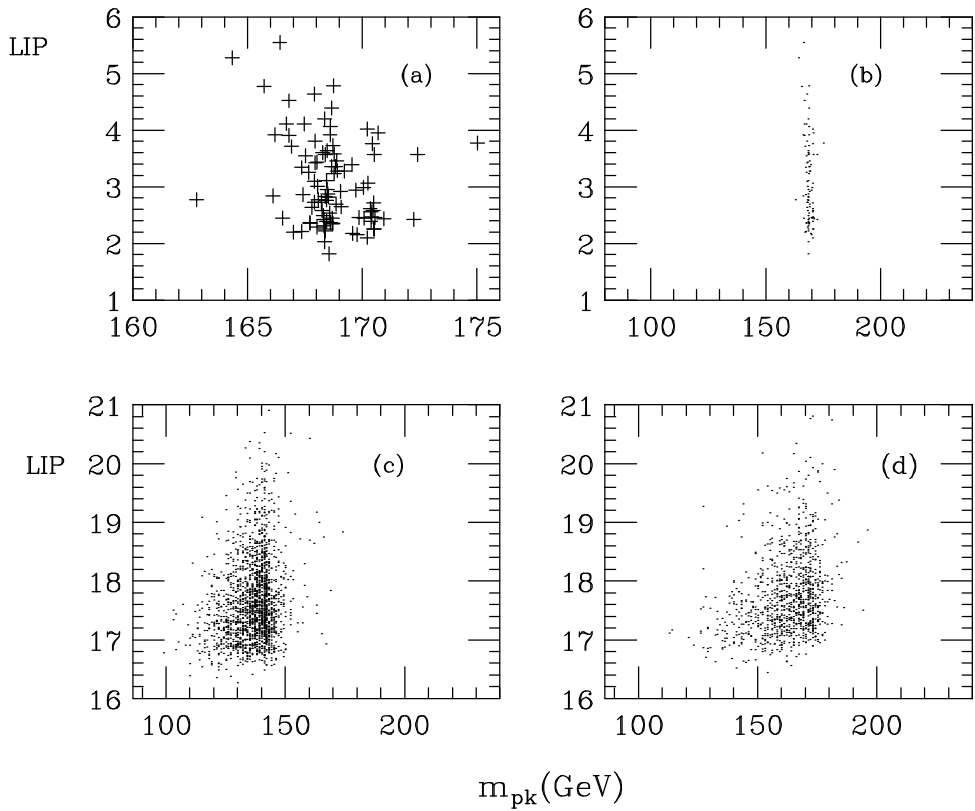}\end{center}
\caption{Scatter plots of $LIP$ vs. peak mass $m_{pk}$ analysed by our
method for Monte Carlo-generated events, as follows: (a) for 95
unileptons, calculated for $m_0=170$ GeV., (b) the same using a smaller 
abscissa, as for the following figures, (c) for 1899 dilepton events,
calculated for
$m_0=140$~GeV., (d) for 1070 dilepton events, calculated for
$m_0=170$~GeV.}
\end{figure}
then made use of these weights to obtain
a smaller set of unweighted events to form a representative subsample.
The largest unilepton subsample we can analyze is about 100, rather
than 1000, since our analysis procedure is quite complicated and takes
more computer time than does the analysis of dilepton events to be
discussed in subsection 2(b).  We assigned a random number $v_i$
between 0 and 1 to each of the chosen events (here 1292) and rejected
those for which $v_i$ exceeds $w_i$ (Barger \& Phillips 1987). 
This left a subsample of 95 unilepton events which was a manageable
batch.  Our purpose then was to compare the observed features of the
candidate top-antitop events with the features predicted for these
events by the QCD model.

Since these were MC events, we know which quarks are which.  In
analyzing each unilepton event, typical energy measurement errors
are assumed for the lepton and the jets but angles are accepted
``as is'', since their measurement is much more accurate than those
for energies.  The jet energies are assumed 
\begin{figure}
\begin{center}\leavevmode\epsffile{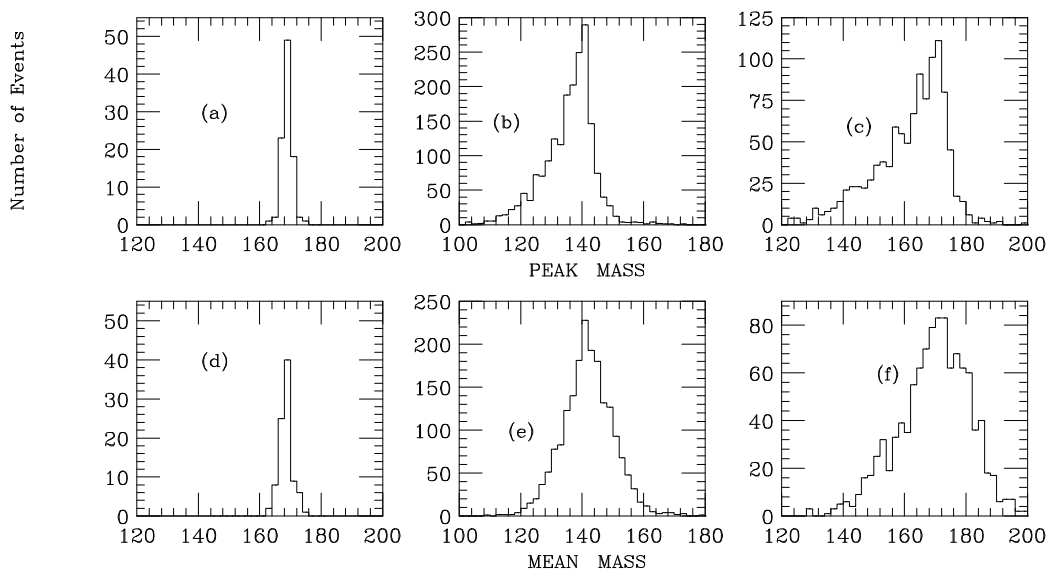}\end{center}
\caption{Projection of the scatter plots of Fig.1 onto the 
$m_{pk}$ axis - (a) for 1(b), (b) for 1(c), the case $m_0=140$~GeV, and
(c) for 1(d), the case for $m_0=170$~GeV. The analogous
projections from the ($\bar{m}(i),LIP(i)$) scatter plots, not shown
here, using mean masses $\bar{m}(i)$: (d), (e) and (f).}
\end{figure}
to have a Gaussian 
distribution in magnitude, which is represented by taking 10 points
about the central value, weighted to give a discrete approximation
to a Gaussian with the known standard deviation $\sigma(E) = 3.4 +0.1 E$
GeV.  We then used our analysis method (Dalitz \& Goldstein 1994,
1992a,b and Goldstein, Sliwa \& Dalitz 1993) to deduce the probability
distribution $P_i(m)$ for each unilepton event, where $m$ denotes the
mass variable.  The parameters $m_{pk}(i)$ for the
peak of $P_i(m)$ and $IP(i)$, the integrated probability given by 
\begin{equation}
IP(i) \equiv \int dm P_i(m), 
\label{IP}
\end{equation}
are convenient for specifying concisely the most 
characteristic features of the probability distribution $P_i(m)$
deduced for an event $i$. In practice we have found it more convenient
to use $LIP$, log$_{10}$ of the integrated probability, defined in
\begin{figure}
\begin{center}\leavevmode\epsffile{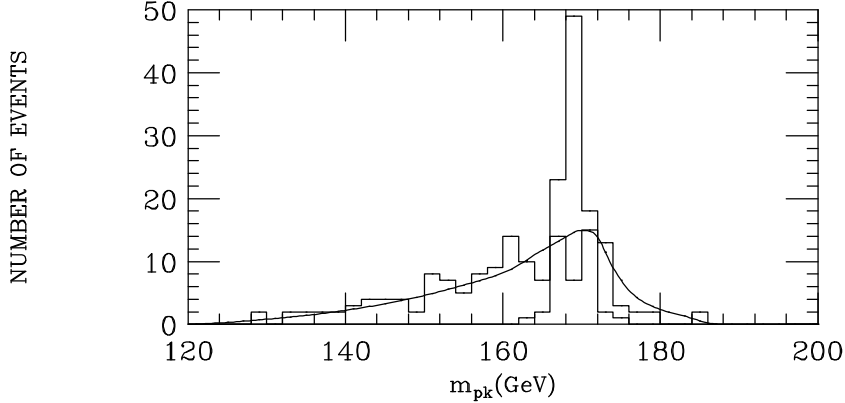}\end{center}
\caption{The distributions of $m_{pk}$ for 153 Monte Carlo dilepton events
(broad histogram), 95 Monte Carlo unilepton events (central histogram),
and for 1070 Monte Carlo dilepton events scaled to 153 events and
smoothed (solid curve).} 
\end{figure}
eq.(\ref{IP}) above,
as our second parameter.  We note again the notation $m_0$ for the
input top quark mass for the Monte Carlo generation of our 
sample of events for testing our procedure.  Other parameters of
relevance are the mean mass $\bar{m}(i)$ for the $i$-th event
\begin{equation}
\bar{m}(i) = {\int m P_i(m) dm \over \int  P_i(m) dm} ,
\label{mean}
\end{equation}
while the mean value of all the $\bar{m}(i)$ in a batch of N events is
\begin{equation}
\langle \bar m \rangle = (\sum_{i=1}^N \bar{m}(i))/N
\label{avemean}
\end{equation}

For our 95 unilepton Monte Carlo events, a scatter plot showing the
distribution of the two parameters $m_{pk}(i)$ and $LIP(i)$, is
displayed in Fig.1(a) and again in 1(b).  This shows that the MC
event-points
are very localized in mass, but widely spread in $LIP$.  This is clear in
Fig.~2(a), which shows the projection
of event-points on the $m_{pk}$ axis (such a distribution of mass values
will be called $\mathcal{P}$$(m_{pk})$ in the following), and is
emphasized
further in Fig.~3,
where the central histogram shows that there is a sharp peak in
$\mathcal{P}$$(m_{pk)}$ at 
approximately 168.4 GeV, only 1.6 GeV below the input mass $m_0 =
170$ GeV, the mass value for which these MC events were computed. The $LIP$
distribution, obtained by projecting the scatter plot on the $LIP$ axis,
is shown in fig.~4(a).
\begin{figure}
\begin{center}\leavevmode\epsffile{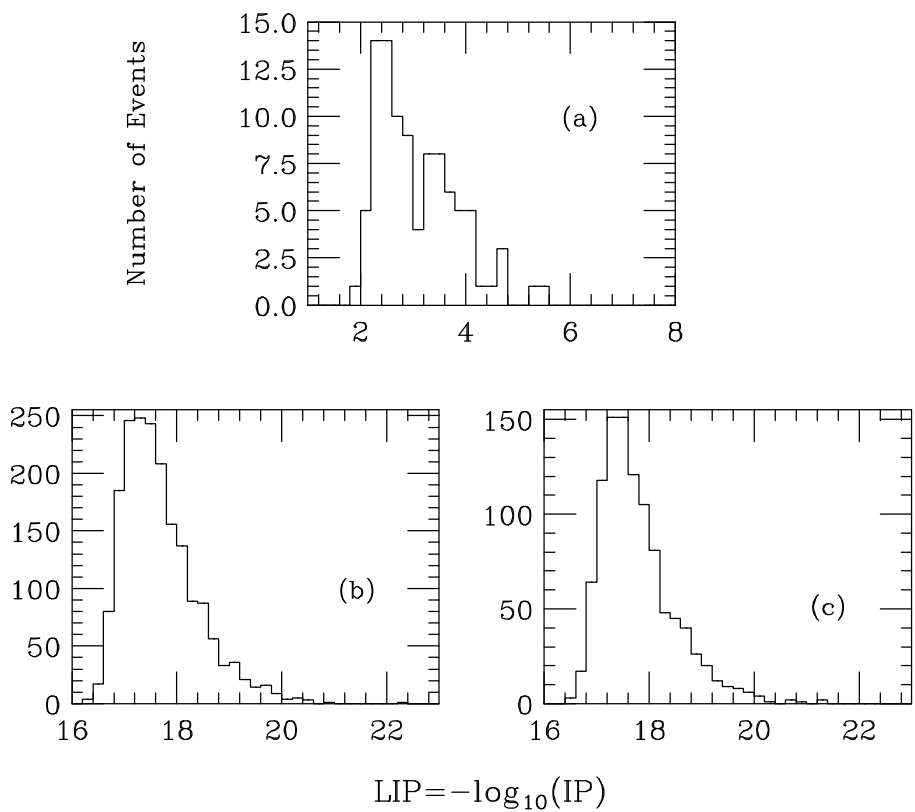}\end{center}
\caption{The projection of the scatter plots onto the $LIP$ axis (a) from
1(a) for 95 unilepton events with $m_0=170$ ~GeV, (b) from 1(c), for 1899
dilepton events with 
$m_0=140$ ~GeV, (c) from 1(d), for 1070 dilepton events with 
$m_0=170$ ~GeV.} 
\end{figure}

The integrated probability values $IP(i)$ given here may appear unreasonably
small, at first sight, but these are the values which result from the
analysis of our Monte Carlo generated events.  To the extent that the
latter are representative of real top-antitop events, these magnitudes
will be typical of the log($IP$) values deduced from real events, and
this will be illustrated by the application of our analysis method to
the real events available, in Sections 3 and 4.  This is the underlying
logic of our method.  If we had a large sample of top-antitop candidates,
which happened to yield $m_{pk}$ values in the mass region of interest
for top, say $\sim 170$ GeV, the evaluation of the
$LIP=-{\mathrm{log}}_{10}(IP)$ for them 
would constitute a test of their interpretation as top-antitop events.
If these $LIP$ values did not follow the $LIP$ distribution shown here in
fig.~4(a), these candidate events could not be accepted as top-antitop
production and decay events.

\subsection{The analysis procedure for dilepton events}

For dilepton events, we used the same randomly chosen sets of eleven
numbers (\ref{angles}) to define 100,000 configurations for the dilepton
events.
Both ($\theta_l^+,\phi_l^+$) and ($\theta_l^{-},\phi_l^{-}$) are for
leptons,
the former for the $l^+$ from the $W^+$ and the latter for the $l^{-}$
from $W^-$ decay.  This change from the case of unilepton events
affects both the kinematics of the event and the cross section
calculated for it using our simple QCD model.  Some of the cuts to be
applied also change when a $q\bar{q}$ pair is replaced by a lepton pair
($l \nu_l$), a further change to the input cross sections.

As in the unilepton case, these 100,000 event configurations were divided
into 5 batches of 20,000 events.  The first batch led to a net weight,
different from that for the unilepton case, in each initial cell.
A second set of partitions of the variables (\ref{angles}) were chosen,
leading
to a second set of cells which are different from the second set for
the unilepton case and which also lead to differing weights in these
cells.  A third set of partitions of the variables (\ref{angles}) has to
be
chosen and so on, to the fifth batch.  We then assigned a random number
$v_i$, between 0 and 1, to each of the 12,503 events surviving 
the cuts in this batch, and rejected those for which $v_i$ exceeds the
weight $w_i$ (Barger \& Phillips 1997). This left us with a random
subsample of 1070 dilepton events for analysis.  In this analysis we
did not include simulated measurement errors, both because the samples
of dilepton events need to be larger, for reasons to be seen below, and
because of time constraints on the computations.  The scatter plot for
the ($m_{pk}, LIP$) values obtained for those 1070 events is shown on
Fig.~1(d).
We note that the mass distribution obtained from the dilepton events 
shown in Fig.~2(c) is very much broader than that shown in Fig.~2(a) for
unilepton events.
This is emphasized again in Fig.~3, where we have
plotted the distributions of peak masses, $\mathcal{P}$$(m_{pk})$, for
both the 95 unilepton events and the dilepton
plot of 1070 events but scaled to 153 events.  The great breadth of 
$\mathcal{P}$$(m_{pk})$ for dileptons -
the FWHM estimated from Fig.~2(c) is about 15 GeV, compared with
about 4 GeV for $\mathcal{P}$$(m_{pk})$ for unileptons - means that many
more dilepton
events must be available for measurement, by about one order of 
magnitude, to gain the same statistical accuracy for the top mass
as with unilepton events.  Balanced against this is the fact that
dilepton candidate events suffer less from background than is the case
for unilepton candidate events.  

The dilepton mass distribution $\mathcal{P}$$(m_{pk})$ also shows very
considerable
asymmetry, in general.  For $m_0 =170$ GeV, the mean peak mass $<m_{pk}>$ 
for 1070 events is 162.0 GeV and the median of the peak mass distribution
is 165.0 GeV, both
well below the peak mass deduced to be approximately 171 GeV.  Indeed, we
find that 74\% of
the dilepton events for $m_0 = 170$ GeV give peak mass values below
$m_0$.  With small samples of dilepton events, it is therefore natural
to find lower top mass values than those from comparable samples of
unilepton events.  We had noted this disconcerting tendency in our earlier
work on the analysis
of dilepton candidate events (Dalitz \& Goldstein 1995).

We have carried out calculations similar to the above, for 100,000
dilepton events with $m_0 = 140$ GeV.  The resulting scatter plot
is given in Fig.~1(c);  the number of events plotted is 1899, and
the mass distribution $\mathcal{P}$$(m_{pk})$ appears much sharper than
for $m_0 = 170$
GeV. For both $m_0 = 140$ and $170$ GeV, the peak obtained for
$\mathcal{P}$$(m_{pk})$ is
in the bin ($m_0, m_0+2$) but its distribution lies mostly below $m_0$,
being markedly asymmetric. The mean mass $<m_{pk}>$ is 136.1 GeV and the
median is 138.0 GeV.

\begin{table}
\caption{Iterations of Monte Carlo simulation for dilepton
events}
\longcaption{Averaging over the batches 6 to 10 gives the more definite
value 4.718 pb for $10^2\cdot\sigma_{net}$, with
$10^2\cdot$fluctuation reduced to 0.018.}
\begin{tabular}{|c|r|r|r|r|r|} \hline
batch no. &  1 &  2 &  3 &  4 &  5 \\ \hline\hline
100$\cdot\sigma_{net}(pb)$ & 4.595 & 4.766 & 4.767 & 4.710 & 4.733 \\
\hline
100$\cdot$fluctuation & 0.133 & 0.059 & 0.045 & 0.040 & 0.039 \\
\hline\hline 
batch no. &  6 &  7 &  8 &  9 & 10  \\ \hline\hline
$100\cdot\sigma_{net}(pb)$ & 4.716 & 4.683 & 4.760 & 4.714 & 4.716 \\
\hline
$100\cdot$fluctuation & 0.039 & 0.039 & 0.039 & 0.040 & 0.040 \\ \hline
\end{tabular}
\end{table}

To investigate the convergence of the iterative procedure used, we
went on to generate another 100,000 dilepton configurations for
$m_0 = 170$ GeV, dividing them into five batches of 20,000 and 
starting from the cells obtained from the fifth batch of the first
100,000 configurations.  The calculations of the total cross section
are extended in Table~1 up to the tenth iteration.  We note that the
fluctuations in the test function (here the total cross section)
from batch to batch are considerable, but that a value accurate to
2\% is achieved after the first five batches; taken together the last 5
batches reduces the accuracy to 1\%. With cuts similar to
these, the total cross section for dileptons has been calculated as
a function of the top mass by several groups (Berends {\textit{et al.}}
1991,
Han \& Parke 1995), using the same QCD tree-level model;  reading
from their graph, their results are about 0.05 pb. for $m_0= 170$
GeV and c.m. energy 1.8 TeV.  Another test function could be provided
by $\mathcal{P}$$(m_{pk})$, the probability function for the top mass.
With this
we found that there was a significant change in $\mathcal{P}$$(m_{pk})$ 
after the first iteration; but that thereafter, $\mathcal{P}$$(m_{pk})$ 
varied moderately 
from batch to batch but without any clear convergence to a limit.

\subsection{Peak Masses or Mean Masses?}

One obvious difficulty about using peak mass values $m_{pk}(i)$ is that
the $P_i(m)$ distributions obtained tend to have spiky and/or multiple
peaks, as is apparent from Fig.~8.  This has the consequence that the 
$m_{pk}(i)$ can be strongly affected by statistical fluctuations,
a rather unsatisfactory situation.  The use of the mean mass $\bar{m}(i)$
does not suffer from this defect, but, of course, it represents a
departure from the use of the Bayes Theorem.  However, it is useful
to trace here the effects its use would have with our Monte Carlo
generated
events.  The event-points scatter more widely when we use the scatter
plot for ($\bar{m}(i), LIP(i)$) rather than ($m_{pk}(i), LIP(i)$),
but the main contrast is in the distribution of $\bar{m}(i)$,
$\mathcal{P}$$(\bar{m})$.
These features can be seen clearly in the $\bar m_i$ mass distribution
shown in Fig.~2(d,e,f).  There is only a slight broadening to be seen for
the
unilepton case, but the dilepton $\mathcal{P}$$(\bar{m})$ for $m_0 = 170$
GeV is
much more symmetric than the $\mathcal{P}$$(m_{pk})$ distribution; its
peak value
is ${(\bar m)}_{pk} = 172$ GeV, its FWHM being about 23 GeV, and its
mean value is $\langle \bar m \rangle =  169.5$ GeV.  The latter
would appear to be quite a reliable indicator for the top quark mass. A
batch of about 25 events, at least in ideal circumstances, should be
sufficient to determine $<\bar{m}>$ to an accuracy of $\pm 2.5$ GeV.
However, to determine $m_t$ from this quantity requires a knowledge of
$\bigtriangleup(m_t) = <\bar{m}(m_t)-m_t>$ as a function of $m_t$, which
can be taken from the analysis of Monte Carlo events discussed above; we
have determined $\bigtriangleup(m_t)$ for the values $m_t=170$ GeV and
(see below) $140$ GeV. This requires some dependence on theoretical
calculations, but at least the dilepton events do have a low background.

The $LIP$ distribution for dileptons,
shown in fig. 4(c), is not unlike that for unileptons in shape.  The
difference in its normalization between these two cases is unimportant;
what matters is that the same normalization should be adopted for the
analysis of real events as for the normalization of our Monte Carlo
events.

As shown in Figs.~1(c), 2(b,e) and 4(b), we have also carried out analyses
for a comparable body of Monte Carlo unilepton and dilepton top-antitop
events for $m_0 = 140$ GeV.  The $m_{pk}(i)$ distribution is
asymmetric, but not as much as for $m_0 = 170$ GeV;  the $\bar{m}(i)$
distribution in Fig.2(e) is symmetric, with peak in the bin (140,
142) and with mean value at 142.0 GeV and median value 141.7 GeV.

The scatter plots display a single mass value for each event, either the
peak or the mean value, as well as the $LIP$. However, the full
distribution $P_i(m)$ for a
single event $i$ contains more information than these two numbers. In
principle, we should 
consider the product of all $P_i(m)$ for a fixed $m$, and use this joint
probability in making deductions from the data. We illustrate this
procedure in Sec.4.

What we have gained from this analysis of Monte Carlo $t\bar t$ events
is some feeling for what variations in $LIP$ and $m$ may arise from
purely statistical origin.  For either dilepton or unilepton events,
the $LIP$ can vary from its mean value within about $\pm 1$.  For the
unilepton events, the peak mass determined will lie very close to the 
top mass, within several GeV below, whereas the mass value determined
from dilepton events has a larger range of variability.

\section{Data and Analyses for Dileptonic Events}

In this Section, we shall discuss what light the Monte Carlo model
calculations above may throw on the interpretation of the dileptonic
events believed to represent top-antitop production and decay in
proton-antiproton collisions with energy 1.8 TeV.  The final states
are
\begin{equation}
{\rm(i)}\quad b \bar b l_1^+ \nu_1 l_2^- \bar\nu_2
\qquad{\rm or}\qquad {\rm(ii)} \quad 
b \bar b l_2^+ \nu_2 l_1^- \bar\nu_1.
\end{equation}
where the leptons ($l_1, l_2$) may be two different species,
($\mu^{\pm},e^{\mp}$),
or the same species, thus ($e^+,e^{-}$) or ($\mu^+,\mu^{-}$).  For the
latter it is
necessary to avoid $e^+ e^-$ and $\mu^+ \mu^-$ pairs arising from the
production and decay of $J/\psi$ mesons, by the use of cuts to exclude
them.

These events (3.1) have two hadronic jets, one generated by the $b$
quark and the other by the $\bar b$ quark, and it is of importance
to identify which jet stems from which quark.  There are a number of
ways to achieve this.  The most direct uses the fact that the $b$
quark has quite a long lifetime $\tau_b \sim 1.6\times 10^{-12}$ sec.
At the high energies in 1.8 TeV collisions, the flight path of the
$b$-system will be of order $5 (E_b/m_b)\times 10^{-2}$ cm, a
macroscopic distance which can frequently be observed with the use
of a specially designed Silicon Vertex Detector.  CDF has had such a
detector in place during its runs over the last two years, referred
to as an SVX, and its effectiveness has been great, as is indicated
by CDF's report that it has efficiency of 40\% for the identification
of $b$ and $\bar b$ decay vertices in the events (3.1).  This is termed
an ``SVX-tag''.  The other instrumental means for determining the 
($b, \bar b$) assignments to the two jets is the observation of the 
charge sign for a secondary lepton arising from the subsequent decay $ b
\rightarrow l^+$ or $ \bar b \rightarrow l^-$.  Since the energy released
in these decays is relatively small, the path of the secondary lepton is
generally close in direction to that of the parent quark jet, while its
charge identifies its quark origin.  This is known as a Secondary Lepton
Tag, SLT for short.  It is a powerfully informative tag but its efficiency
is lower, being about 20\%,  as CDF reports.

Other means exist for deciding which is the $b$-quark, from internal
evidence about the event, in its analysis.  For example, our analysis
procedure involves taking the lepton $l_1^+$ together with a jet, say j1, 
and determining the spatial circle $E_{1+}$ on which the total momentum
(${\bf P}_{j1} + {\bf l}_1^+$) must lie if its parent $t$ quark mass
is $m_1$, and then taking the lepton $l_2^-$ with the other jet and 
determining the spatial circle $E_{2-}$ on which the total momentum
${\bf P}_{j2} + {\bf l}_2^-$ must lie if their parent $t$-quark mass is
$m_2$.  If this event really is ($t + \bar t$) decay, then $m_1 = m_2 
=m$.  Further, the momentum $( {\bf t} + \bar{\bf t} )_T$ transverse
to the initial proton-antiproton axis is necessarily very limited in
magnitude and might even 
be constrained to be zero, in first approximation, in which case the 
possible transverse momenta $({\bf P}_{j1} + {\bf l}_1^+)_T$ and 
$({\bf P}_{j2} + {\bf l}_2^-)_T$ are given by the intersections of
the two ellipses obtained by projecting the spatial circles
$E_{1+}(m)$ and $E_{2-}(m)$ onto the transverse plane. These two
projected ellipses may have 4, 2 or 0 intersections, depending 
\textit{inter alia} on the the value of $m$.  If they have no intersection
for any $m$, this just means that this pairing of the leptons with
the jets is not consistent with a $t$-$\bar t$ origin.  However, it is
also possible to associate the pairs $({\bf P}_{j2} + {\bf l}_1^+)$
and $({\bf P}_{j1} + {\bf l}_2^-)$ instead, to repeat these steps with
them, and to find the resulting ellipses do have sensible intersections.
If this is the case, we can conclude that $j_2$ is the $b$-jet and that
$j_1$ is the $\bar b$-jet.  This is the case for the event 19250,
for example.

Since there are generally ranges of $m$ for which the projected ellipses
have two intersections (and sometimes even ranges where there are four
intersections, although this is  rare), we can take this line of
argument a step further.  For each corresponding solution as a 
function of mass $m$, there will be a value for $LIP$.  If there
is a solution for which $|LIP|$ is much lower than all other solutions,
it is natural, following the Bayesian principle, to confine attention
to this solution. This choice represents a definite assignment of the 
jets ($j_1, j_2$) to the quarks ($b, \bar b$).

\begin{table}
\caption{Dilepton event data reported by CDF (Abe, \textit{et al.} 1990,
1994b; Sliwa 1991)}
\longcaption{Comments on the right are from CDF. Elsewhere in the text we
shall identify each event by its run number only; events are so rare
that, to date, identification by run number has been unique.} 
\begin{tabular}{|c|r|r|r|r|l|} \hline
       &   $p_x$ &  $p_y$&   $p_z$&$E(GeV)$&   \\ \hline\hline
\multicolumn{6}{l}{Run 19250  Event 20425}     \\ \hline
$e^+$  & -21.2 & 23.6 & -28.6 &  42.7 &   \\
$\mu^-$& -0.6  &-43.7 & -38.6 &  58.3 &   \\
jet 1  &  18.7 & -6.3 &  25.3 &  33.3 &   \\
jet 2  &  0.7  & 8.9  & -70.1 &  70.7 &   \\
$\mu^+$& -1.0  & 7.9  & -28.7 &  29.8 &   \\ \hline
\multicolumn{6}{l}{Run 41540  Event 127085}    \\ \hline
$e^-$  &  18.7 & 11.7 &  20.7 &  30.2 &   \\
$\mu^+$&  46.1 & 11.5 &   8.1 &  48.2 &   \\
$\mu^+$&   8.7 & -1.2 &   1.6 &   8.9 & soft $\mu^+$  \\
jet 1  & 129.7 &-18.2 &  14.4 & 131.8 & with $\mu^+$  \\
jet 2  & -50.0 &-35.0 & -34.6 &  70.1 & with $e^-$    \\
jet 3  &  -9.7 & 24.1 &-245.2 & 246.6 & backward anti-proton jet \\
                                               \hline
\multicolumn{6}{l}{Run 47122  Event 38382}     \\ \hline
$e^+$  &  45.9 & 21.4 &  54.1 &  74.1 &   \\
$\mu^-$&  37.2 &  2.6 & -30.2 &  48.0 &   \\
jet 1  & -67.0 &-52.3 &  58.2 & 103.0 &  with $e^+$ \\
jet 2  &  25.0 & -7.2 &  46.2 &  53.1 &  with $\mu^-$ \\
jet 3  &  17.3 & -5.0 &-246.1 & 246.8 &  backward anti-proton jet \\
                                                \hline
\end{tabular}
\end{table}

We give on Table 2 the lepton and jet energies and momenta for the three
early CDF events, for illustration.  The event 47122 is a normal
event.  The jet 1 is associated with $e^+$, so that it is the $b$-jet;
the jet 2 is associated with $\mu^-$, and it is the $\bar b$-jet.
The jet 3 has an exceedingly large longitudinal momentum with a modest
transverse component $|P_{3T}| = 18$ GeV, making an angle of 4
degrees with the initial $p\bar{p}$ axis, and belongs to the
remnants of the antiproton after the collision. The event
19250 has a secondary $\mu^+$ lepton of energy 30 GeV which is
clearly associated with the $\bar b$-jet of energy 71 GeV, its
transverse momentum to the $\bar b$-jet being only 4 GeV.  The third
event, 41540, also includes a third jet of energy 247 GeV
which makes only a 6 degree angle with the initial $p\bar{p}$
direction and clearly belongs to the antiproton remnants after the
collision. Its second $\mu^+$ meson is anomalous, since it is not
associated with jet-2 and we shall comment further on this event
in the Appendix.
\begin{figure}
\begin{center}\leavevmode\epsffile{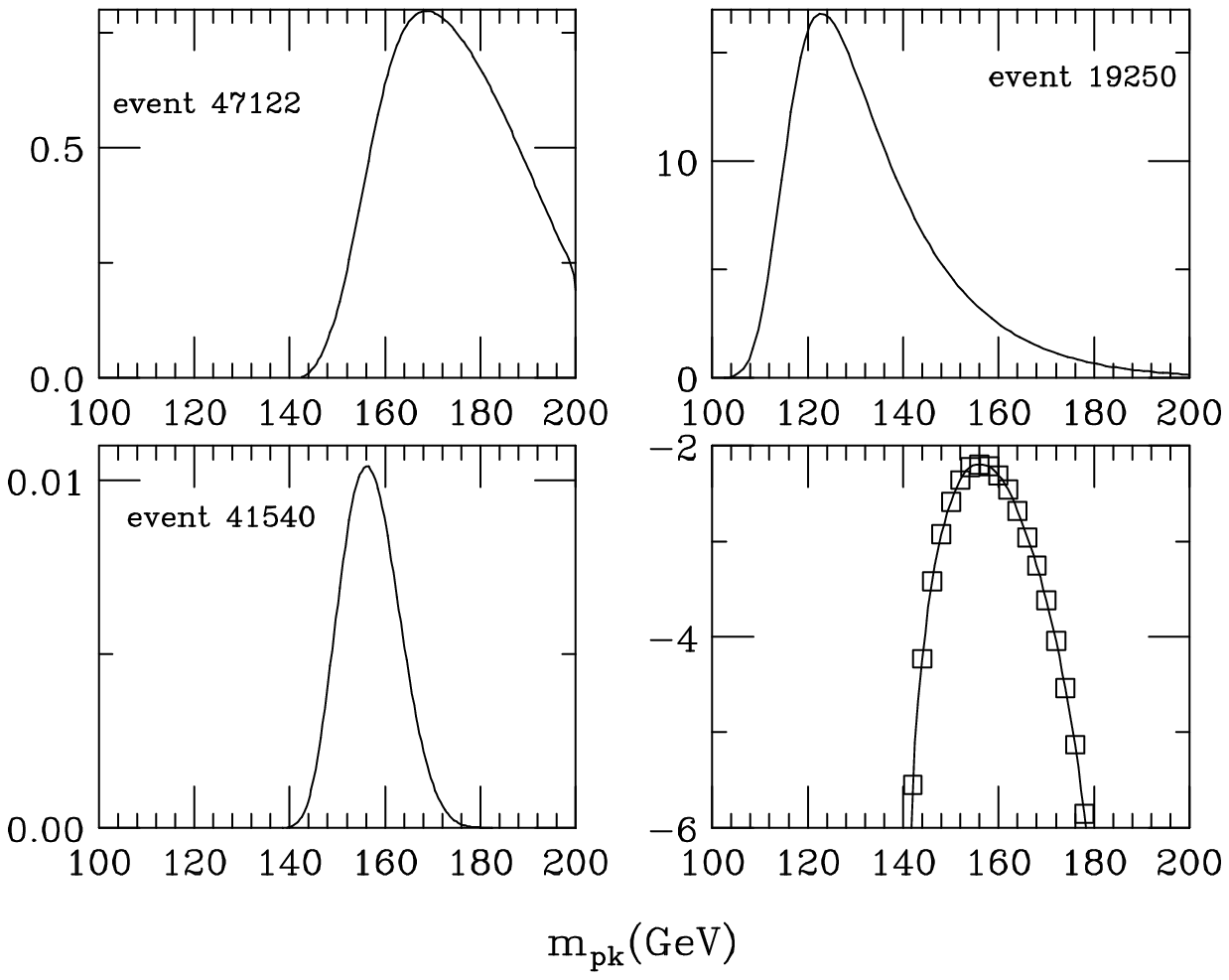}\end{center}
\caption{$P_i(m)$ for three CDF dilepton events (note that we use 
CDF's run number to label the event). The lower right figure is
the logarithm of the product of the three probabilities.}
\end{figure}

 We have analysed these
three events by the same procedure as outlined in the Monte Carlo
section (Dalitz \& Goldstein 1992a,b,1994).
We assigned uncertainties to the jet transverse
energies by the same algorithm as above. The resulting probability  
distributions are shown in Fig. 5.
In each case, the assignment of jets to the $b$ and $\bar{b}$
quarks is uniquely determined by this analysis. 
The probability distributions $P(m)$ from these 
CDF events, shown in Fig.5, peak at $158~{\rm GeV}$, with $LIP=20.0$ 
for 41540, at $168~{\rm GeV}$, with $LIP=17.4$ for
47122 and at the low mass of $121~{\rm GeV}$, with $LIP=16.8$, for
19250. The joint probability for these three events is the product of
their separate probabilities and log$_{10}$(joint probability) for them
is plotted vs. m as Fig.5(d), peaking at about 158 ~GeV. These three
early events illuminated well the tendency for small sets of dilepton
events to indicate lower $m_t$ values than the unilepton events, using
our Bayesian method.

At the 1997 Electron-Photon Conference at Hamburg, CDF reported on
9 dilepton events which included the two events previously reported--7 
$e^{\pm}\mu^{\mp}$, 1 $e^+e^{-}$ and 1 $\mu^+\mu^{-}$--and a
preliminary report on 
their details is given in a Ph.D dissertation (Kruse 1996).
D0 reported on 5 dilepton events--3 $e^{\pm}\mu^{\mp}$, 1 $e^+e^{-}$  and
1 $\mu^+\mu^{-}$--and a preliminary report on their details
is given in a Ph.D. dissertation (Varnes 1996), available on the 
Internet.  We have made our own analysis of these events, with results
quite similar to those reported recently, as just mentioned.  We
give a plot of the $m_{pk}$ values for 15 dilepton events in Fig.~6.
They are distributed very broadly, which we had anticipated following
our analysis of MC-generated QCD events in Sec.~2 above.  This supports
our conclusion that small samples of dilepton events are bound to give
mass values substantially below the top quark mass.  It is noticeable
that the mass values reported by D0 lie systematically above those
reported by CDF;  of 5 D0 events, 3 lie above all of the CDF values,
and one of these is very high, isolated at 200 GeV.  The median
mass for this plot lies in the 155-160 GeV bin and this does not
change if we cut off its extreme values,   two in the 120-125 bin
and three at masses 185, 185 and 200 GeV. We note only that the
mass plot is in good qualitative accord with expectation. 
Unfortunately, we do not yet have Monte-Carlo calculations for 
$m_0 > 170$ GeV but expect that the curve for $m_0 = 180$ GeV will
be close to that obtained by expanding elastically the plots for 
$m_0 = 170$ GeV in the variable $m$.

\begin{figure}
\begin{center}\leavevmode\epsffile{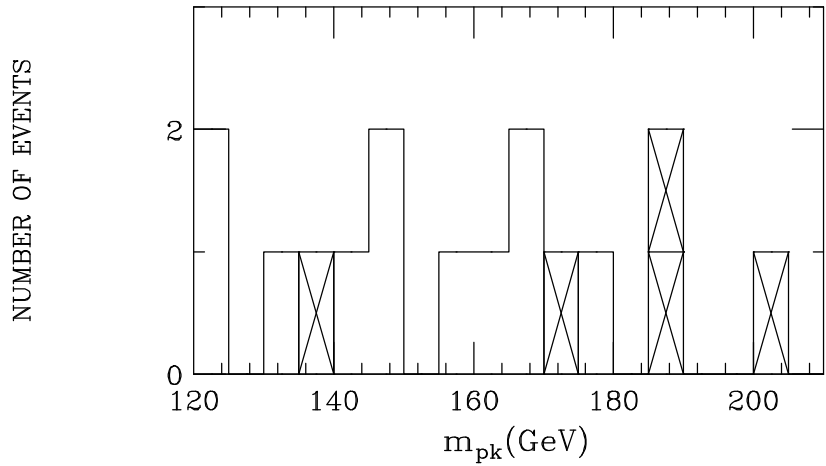}\end{center}
\caption{The distribution of $m_{pk}$ values determined from the 16
dilepton events available empirically. The D0 events are marked with an
X.}
\end{figure}

Finally, we turn to the projection of the scatter plot onto
the $LIP$ axis, shown for the CDF and D0 data on Fig.~7.
This is to be compared with the distributions in Fig.~4 which
we have calculated for the values $m_0 = 140$ and $170$ GeV.
For both of these, the peak in $LIP$ was found to be about
$LIP$ = 17.2, the rapid rise from below being at about 
$LIP$ = 16.8.  The empirical values for $LIP$ are of the right
order-of-magnitude but appear to be smaller than the calculated
values by about 0.8.

\begin{figure}
\begin{center}\leavevmode\epsffile{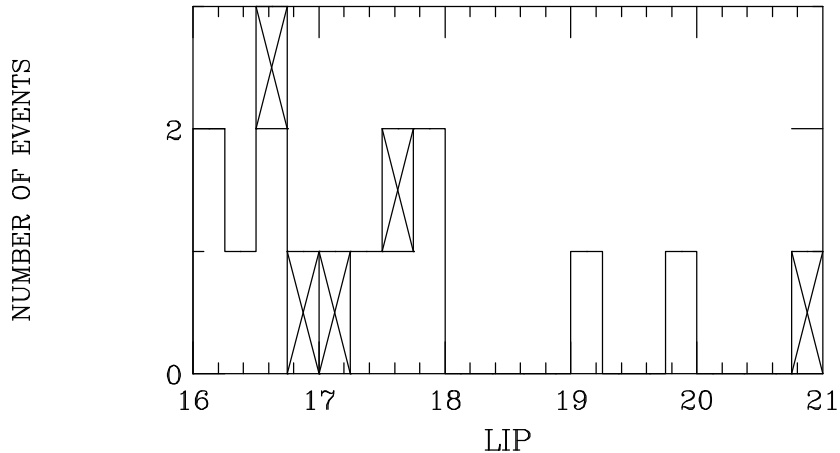}\end{center}
\caption{The distribution of $LIP$ values determined from the 16 dilepton
events available empirically. The D0 events are marked with an X.}
\end{figure}

\section{Data and Analyses for Unilepton Events}

In this Section, we discuss the available data on the final states
\begin{equation}
{\rm(a)} \quad b(l^+ \nu_l) \bar b(\bar q q), \qquad\qquad
{\rm(b)} \quad b(\bar q q) \bar b(l^- \bar \nu_l)
\end{equation}
which result when one $W$-decay is leptonic, for $l = e$ or $\mu$,
and the other $W$-decay is to two hadron jets, in the light
of our model calculations described in Sec.~2 above.  The states
(4.1) have the advantage that there is only one neutrino, but they
also have the disadvantage that there are four jets to be identified.
It is clear that $b$-tagging is the vital key to their unique, or
most probable identification.

The data on unilepton events reported at the lepton-photon conference
at Hamburg (Geromini 1997) are now quite large in number.
The CDF collaboration reported 22 $b$- and/or $\bar b$- tagged events,
made up of 12 events with a single SVX-tag, 8 with a single SLT-tag
and 2 with a double tag.  This includes the 7 tagged events 
reported earlier (Abe {\textit{et al.}} 1994b), when their SVX detector
was already in use.  With allowance for background, CDF have used
their events to give an estimate of the top mass, namely $m_t =
176.8(4.4)$ GeV.  The D0 collaboration reported on 11 events, 5 having
a primary electron and 6 having a primary muon, on the basis of which
they made the estimate $m_t = 173.3^{+5.6}_{-6.2}$~GeV.
Neither CDF nor DO reported on individual unilepton events beyond
CDF's 7 $b$-tagged events just mentioned, so we are necessarily limited
to the discussion of the latter.

\begin{table}
\caption{Single lepton data (Abe {\em et al.} 1994b)}
\longcaption{In each event (identified by run number) one jet was tagged
as a $b$-jet via
the Silicon Vertex Detector (SVX) or the emission of a Soft Lepton (SLT).
The $j1,...$ labels correspond to CDF's jet numbers.}
{\small
\begin{tabular}{|c|c|r|r|r|r|l|} \hline
event  &     &   $p_x$ &  $p_y$&   $p_z$&$E(GeV)$&   \\ \hline\hline
40758& $e^+$  & -94.313 &-50.113&  48.523& 117.306&    \\
& jet  j1&  86.267 & 26.685& -21.881&  92.913&  SVX \\
&     j2& -26.220 & 74.310&  23.996&  82.373&  \\
&     j3&  46.052 & 47.417&  43.659&  79.217&  \\
&     j4&  30.613 &-22.003&  76.790&  85.545&  \\ \hline
43096 & $e^-$  &  21.753 & 21.093& -27.316&  40.795&   \\
   &  j1&  78.068 &100.425&   2.544& 127.225&  SVX  \\
   &  j2& -70.137 & 29.785& 137.091& 156.845&  \\
   &  j3&  10.642 &-66.960&  81.787& 106.235&  \\
   &  j4& -34.707 &-14.202& 138.856& 143.831&  \\ \hline
43351 & $\mu^-$&  24.577 & -1.062&  -1.723&  24.660&  \\
   &  j1& 109.365 &-75.333& 195.687& 236.494&  \\
   &  j2& -85.879 &  6.159& -15.582&  87.499&  \\
   &  j3&  24.815 &-21.905&   7.680&  33.979&  \\
   &  j4&  -3.595 & 36.122&  14.128&  38.953&  SLT \\
   & $\mu^-$&  -0.605 &  2.032&   1.057&   2.369&  $p_{l\perp}=0.46$ \\
\hline
45610 & $\mu^+$&  52.325 & 11.153&  -9.682&  54.369&  \\
   &  j1&  11.612 & 76.423& -58.639&  97.025&  SVX  \\
   &  j2& -13.843 &-71.064& -74.320& 103.755&  \\
   &  j3&   3.167 &-36.061& -77.935&  85.932&  \\
   &  j4& -19.286 & -9.042&   1.492&  21.352&  \\ \hline
45705 & $e^-$  &  12.221 & 54.445&  42.329&  70.038&  \\
   &  j1& -74.864 &-49.953&  81.137& 121.174&  \\
   &  j2& -51.229 &  4.189& -11.399&  52.649&  \\
   &  j3&  15.072 &-54.971&  41.817&  70.694&  SLT \\
   &  j4&  31.974 & -9.305& 113.471& 118.256&  \\
   & $e^+$  &   1.523 &-10.995&   8.984&  14.280&  $p_{l\perp}=1.62$ \\
\hline
45879 & $\mu^+$&  52.586 &  4.746& -11.170&  53.969&  \\
   &  j1& -75.575 & 27.724&-197.545& 213.317&  \\
   &  j2&  45.871 &-84.446& -10.592&  96.682&  SVX \& SLT \\
   &  j3&  33.259 & 25.812&   5.488&  42.456&  \\
   &  j4& -36.642 & -4.359& -16.765&  40.530&  \\
   & $\mu^-$&   6.680 &-11.732&  -1.488&  13.582& $p_{l\perp}=0.27$ \\
\hline
45880 & $e^-$  &  -5.942 &-25.106&   4.146&  26.131&  \\
   &  j1&  98.037 & -7.188& -19.791& 100.273&  \\
   &  j2& -26.071 &-55.037&  80.329& 100.804&  \\
   &  j3&  18.082 & 39.344&  16.853&  46.464&  SLT  \\
   &  j4& -22.051 &-13.776& -16.246&  30.658&  \\
   & $e^-$  &   0.838 &  2.440&   1.088&   2.800& $p_{l\perp}=0.27$ \\
\hline
\end{tabular}}
\end{table}

To illustrate the above remarks about unilepton event analyses, we apply
our procedure to the available
data, the 7 such events published in full by CDF(Abe,
\textit{et al.} 1994). In Table 3, we lay out these data, in a form
convenient for use. All 7 events are b-tagged, three are SLT-tagged, three
are SVX-tagged, and one is both SLT- and SVX-tagged.

The jet calorimeter energies are the ``corrected values'' quoted by
CDF (Abe {\textit{et al.}} 1994b) in their Appendix A, following the
calculated
scatter plots given in their Fig.~57;  they are the CDF estimates for 
the original parton energies, with well defined statistical uncertainties.
The c.m. energy for the two jets hypothesized to result from $W$ decay
is generally rather far from the well-known value (Particle Data Group
1996) $M_W = 80.2(2)$ GeV, and this presents a problem.  CDF uses
a kinematic fitting procedure, established long ago (Dahl {\textit{et
al.}}
1968) in bubble chamber work, to manipulate the already corrected
transverse energies in order to reproduce the $W$ mass value at the
expense of a higher $\chi^2$.  We do not use the resulting ``best fit
values'' given by CDF, since we follow a different scheme of analysis
(Dalitz \& Goldstein 1994).

Our analysis of these $l^\pm 4j$ events (4.1) employs a simple extension
(Goldstein {\textit{et al.}} 1993) of the method used for dilepton events.
We
sketch it briefly here, for the case of a positively charged lepton
$l^+$; the case for $l^-$ follows when every particle is replaced by
its antiparticle and vice versa.  One jet chosen tentatively to be the
$b_l$ jet associated (using the CDF notation (Abe {\textit{et al.}} 
1994b)) with
$l^+$ and a kinematic paraboloid is formed, as before, leading to an 
ellipse in momentum space which includes all momenta $\mathbf{t}$
consistent
with $\mathbf{b_l}$ and $\mathbf{l^+}$, for an assumed mass $m_t$.  The
other three jets
are assumed to arise from $\bar t$ decay where the resulting $W^-$
boson decays hadronically, thus:
\begin{equation}
\bar t \rightarrow \bar b + W^-,\quad {\rm followed \: by }
\quad W^- \rightarrow \bar q_1 + q_2.
\end{equation}
The quark assumed to be $\bar b$ will be denoted by $b_j$. The experimental
error distributions for these quark energies have been discussed in much
detail by CDF (Abe {\textit{et al.}} 1994b) and we adopt the same
algorithm that
interpolates their $\sigma_E$ values as stated in Sec.~2.  A grid of
momentum values ($\mathbf{\bar b}$, $\mathbf{\bar q_1}$, $\mathbf{q_2}$)
is laid out and weighted
by their probability values at each point, together with a probability
weighting $F_W(\bar q_1,q_2)$ of Breit-Wigner form to emphasize those
grid points at which $(\bar q_1,q_2)$ is consistent with the $W$-boson
mass.  At each grid point, there is a definite momentum ($\mathbf{\bar t} 
= \mathbf{\bar b} + \mathbf{\bar q}_1 + \mathbf{q}_2$) and deduced mass
$m$, and this point is
then paired with the points on the $t$-ellipse for $m$, which also have
their weighting factors due to measurement errors.  This product of 
probabilities is finally weighted by a Gaussian factor $G[\mid\! 
({\mathbf{t}}+{\mathbf{\bar t})_T}\!\mid/ \rho]$ to represent the effect
of limited transverse momentum due to
initial state gluon emission, the value 0.1$m$ being adopted for the
parameter $\rho$.  Contributions to the net probability for the top
quark mass to lie within ($m, m+\Delta m$) come from all grid points which
lie within the band $\Delta m$, and are summed to give the net 
probability $P_i(m)$ indicated by this event.

For a real event having a primary lepton $l^+$, the probability must be
computed separately for each possible assignment 
$(b_l;\bar{q}(1),q(2),b_j)$ to the four outgoing quark jets labelled
arbitrarily as j1,j2,j3 and j4. Here $b_l$
denotes the jet arising from the decay $t\rightarrow bW^+ \rightarrow
b_l(l^+\nu_l)$, while $\bar{b}_j$ denotes the jet arising from the decay
$\bar{t}\rightarrow \bar{b}W^- \rightarrow \bar{b}_j(\bar{q}(1)q(2))$
where the $q(i)$ are the light quark jets $(\bar{u}(1)d(2))$ and
$(\bar{c}(1)s(2))$. At the Tevatron, the primary lepton will generally
have high energy, typically of order 40 GeV, and a large $p_T$ with
respect to each of the four jets. For a real event with a primary lepton
$l^-$, the assignments are $(\bar{b}_l;q(1),\bar{q}(2),b_j)$, the quarks
for the $l^+$ case being replaced by antiquarks and vice versa. If the
primary lepton charge is known, as is usual, and none of the jets are
b-tagged, there will be 24 permutations to consider for the identities of
the four jets, with 24 different probabilities, in general. Most of these
24 possibilities will not provide a fit to the event, or at most a very
poor fit; the constraint that the pair $(\bar{q}(1)q(2))$ should have mass
$M_W$ is particularly effective in this respect.

The $b$ and $\bar{b}$ jets can be identified by a secondary lepton tag
(SLT), since $b\rightarrow cW^- \rightarrow cl_2^-\bar{\nu}_l$ while
$\bar{b} \rightarrow \bar{c}W^+ \rightarrow \bar{c}l_2^+\nu_l$. The
secondary lepton $l^{\pm}$ will typically have a modest energy, say from 2
to 15 GeV for Tevatron events, and, most significant, a low $p_T$, say 0.2
to 1.5 GeV/c, relative to one of the jets, since the energy release for
$b\rightarrow c$ is about 3 GeV. For secondary lepton $l_2^+$, the
associated jet will be a $\bar{b}$-jet; for secondary lepton $l_2^-$, the
associated jet will be the $b$-jet. Since $(W^+\rightarrow e^+ \;
{\mathrm or} \; \mu^+)/{\mathrm all}\; W^+ \; {\mathrm decays}) =
21.2(7)\%$, SLT occurs quite often. In the decay of
$(\bar{t}t)$ systems, a single SLT must occur in 34\% of the events, while
a double-SLT will occur in 5\% of the events. In practice, the observed
rates will be lower than these since the detector efficiences must also
be taken into account; the overall efficiency reported by CDF is 20\% for
SLTs. However, with SLT, only 6 permutations need be considered for the
other three jets.

The $b$ and $\bar{b}$ jets have also been identified by CDF(Abe,
\textit{et
al.} 1994b) using their SVX, in which a visible
decay vertex for $b\rightarrow c$ (or $\bar{b}\rightarrow \bar{c}$) has a
large energy release, about 3 GeV. This contrasts with the case of a
c-quark, which has a shorter lifetime, about $0.4\times 10^{-12}$ sec for
the
$D^0$ and $D_s^0$ states and about $1.06\times 10^{-12}$ sec for
$D^{\pm}$, and
whose dominant transition, $c\rightarrow s$ releases much less energy, of
order only 1 GeV. However, the decay vertex does not distinguish between
$b$ and $\bar{b}$-quark, so that a SVX-tag requires that the analysis must
be carried through for both assignments, each with 6 permutations. CDF has
reported that the SVX detector has an efficiency of 40\%, twice that of
the SLT. Multiple $b$-tags and $\bar{b}$-tags will occur; CDF(Abe,
\textit{et
al.} 1994b) has already reported two events with double-SVX tags and one
event with both SVX- and SLT-tags.

It is important to emphasize that each jet should be treated in an
identical way, so that relative probabilities between different events
and/or different interpretations can be meaningfully compared.

\begin{table}
\caption{Output from our analysis of 7 single lepton events}
\longcaption{CDF's fitted
values(Abe {\em et al.} 1994b) using a kinematical program (Dahl,
\textit{et al.} 1968)  are listed below, together with their preferred
jet assignment.}
\begin{tabular}{|c|c|r|c|r|r|l|} \hline
jets               & LIP               &$m_t$            & Best Fit &
             &      CDF $m_t$&CDF $\chi^2$       \\ \hline
($b_l,q_1,q_2,b_j$)&                   &                 &($x,\bar{x}$) &
$m(t\bar{t})$&               &                   \\ \hline
                   &\multicolumn{4}{l}{CDF fitted values appear below}
             &               &                   \\ \hline
\multicolumn{7}{l}{ Run 40758 Event 44414} \\ \hline
(j4,j2,j3,j1)      & 3.4 &$170^{+13}_{-9}$ &(0.395,0.201) &
507          &               &                   \\
                   &                   &                 &(0.481,0.177) &
526          & $172\pm{11}$  & $<$0.1             \\
(j2,j3,j4,j1)      & 5.4 &$184$            &(0.363,0.211) &
498          &               &                   \\ \hline
\multicolumn{7}{l}{ Run 43096 Event 47223} \\ \hline
(j1,j3,j4,j2)      & 5.5 &$162^{+8}_{-4}$  &(0.521,0.139) &
484          &               &                   \\
                   &                   &                 &(0.529,0.152) &
511          & $166\pm{11}$  & 2.0               \\
(j4,j2,j3,j1)      & 7.3 &$224$            &(0.417,0.172) &
481          &               &                   \\ \hline
\multicolumn{7}{l}{ Run 43351 Event 266429} \\ \hline
(j2,j1,j3,j4)      & 6.3 &$160^{+12}_{-6}$ &(0.447,0.168) &
493          &               &                   \\
                   &                   &                 &(0.409,0.157) &
455          & $158\pm{18}$  & 6.1               \\ \hline
\multicolumn{7}{l}{ Run 45610 Event 139604} \\ \hline
(j2,j3,j4,j1)      & 3.8 &$180^{+7}_{-13}$ &(0.110,0.379) &
367          &               &                   \\
                   &                   &                 &(0.093,0.446) &
365          & $180\pm{9}$   & 5.0               \\
(j1,j3,j4,j2)      & 5.1 &$112$            &(0.107,0.396) &
369          &               &                   \\ \hline
\multicolumn{7}{l}{ Run 45705 Event 54765} \\ \hline
(j3,j1,j2,j4)      & 3.4 &$190^{+11}_{-14}$&(0.409,0.148) &
443          &               &                    \\
                   &                   &                 &(0.593,0.097) &
430          & $188\pm{19}$  & 0.4                \\
(j4,j1,j2,j3)      & 4.95 &$156$            &(0.388,0.135) &
411          &               &                    \\ \hline
\multicolumn{7}{l}{ Run 45879 Event 123158} \\ \hline
(j2,j3,j4,j1)      & 3.4 &$179^{+12}_{-10}$&(0.140,0.406) &
423          &               &                    \\
(j2,j1,j4,j3)      & 3.6 &$180$            &(0.131,0.452) &
438          &               &                    \\
(j1,j3,j4,j2)      & 5.7 &$168$            &(0.130,0.273) &
396          &               &                    \\
                   &                   &                 &(0.129,0.420) &
419          &$169\pm{10}$   & 2.2                \\ \hline
\multicolumn{7}{l}{ Run 45880 Event 31838} \\ \hline
(j1,j2,j4,j3)      & 3.4 &$164^{+12}_{-10}$&(0.225,0.176) &
358          &               &                    \\
(j2,j1,j4,j3)      & 3.5 &$134^{+6}_{-8}$  &(0.295,0.358) &
358          &               &                    \\
                   &                   &                 &(0.312,0.132) &
365          & $132\pm8$     & 1.7                \\ \hline
\end{tabular}
\end{table}

	Our analyses of these events are summarized in Table 4, giving  
the values found for $m_{pk}$ and $LIP$ for all of the assignments which  
provide a fit to each event, for both possibilities, (a) that the tagged  
jet in the event is due to a $b$-quark, and (b) that the tagged jet is due
to a $\bar{b}$-quark. This meant ignoring the
charge sign for the secondary lepton for the SLT events; the implications
of this sign are considered in the brief discussion we give below for each
event. In most cases, there is one assignment which is more strongly
favoured than the others; we then confine attention here to the fits,
whether with $b$ or $\bar{b}$, having the lowest $LIP$. Two exceptions to
this remark are events 45880 and 45879, for which there are two fits and
three fits, respectively, having comparable $LIP$ values. The
corresponding probability functions $P_i(m)$ for all of these fits are
plotted in Fig.8.
Their spikiness is due to statistical fluctuations in our numerical
evaluations of the complicated integrals involved; these curves could do
with some smoothing but we have preferred to leave them as they have come
out. 

\begin{figure}
\begin{center}\leavevmode\epsffile{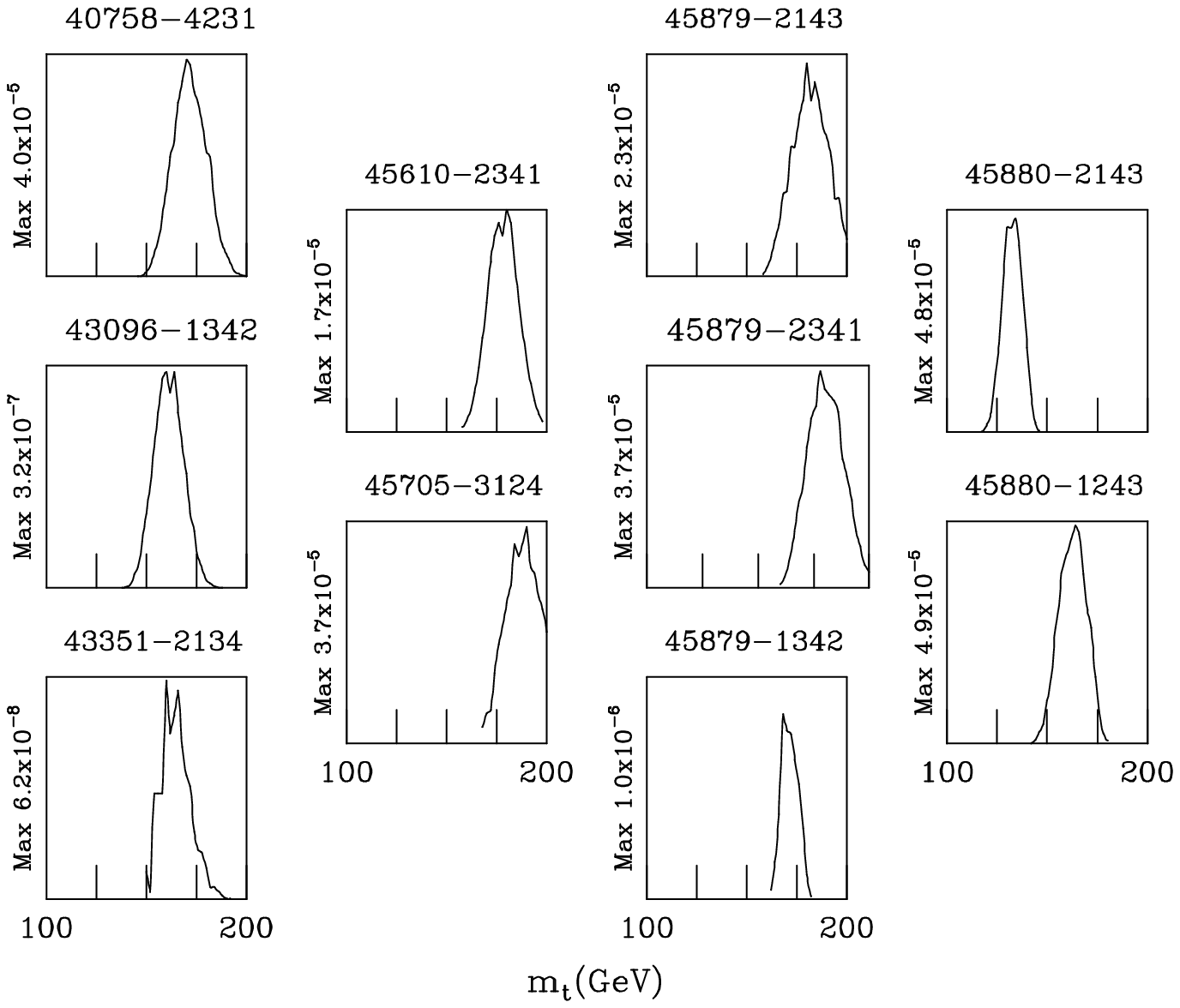}\end{center}
\caption{Plots of $P_i(m)$ for 7 CDF unilepton events, for all promising
jet assignments.}
\end{figure}

	Four of the events have $LIP\simeq 3.4$, the others having the
values 3.8, 5.5 and 6.3. The calculated $LIP$ values for our 95 MC events
peak at about 2.5, rising rapidly from essentially zero at $LIP=2.0$ to a
peak value at $LIP\simeq 2.5$ and then falling by a factor of about 10 by
$LIP=4.5$. The shape of the observed distribution for $LIP$ appears
qualitatively correct but with the $LIP$ scale displaced upwards by about
one unit in $LIP$. 

	The product of the seven independent $P_i(m)$ distributions is  
plotted on Fig.9, where we also show the product calculated without the  
events 43096 and 43351, whose 
$LIP$s exceed 5.0. These curves peak at  
$172_{-4}^{+2}$ GeV, in good accord with the CDF conclusion for these
events. Our general expectation that the distribution of $m_{pk}(i)$  
should consist of a fairly narrow peak appears to be the case.

\begin{figure}
\begin{center}\leavevmode\epsffile{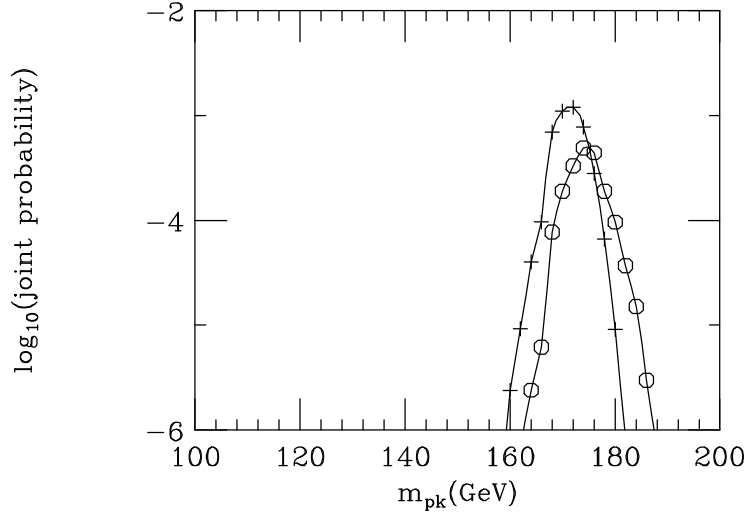}\end{center}
\caption{Logarithm of joint probability for the CDF unilepton events in
Fig.8. The upper curve is for all 7 events; the lower curve has events
43096 and 43351 excluded.}
\end{figure}

	We now turn to a brief discussion of the individual
$l^{\pm}\bar{b}b\bar{q}q$ events. The numbering of the jets, jN for N=1
to 4 for each event, is taken from CDF and given here in Table 3; the
tags are specified by jet number and by type, SVX and SLT. We use the
following notation for the events: $(b_l;\bar{q},q,\bar{b}_j)^+$ and
$(\bar{b}_l;\bar{q},q,b_j)^-$. $b_l$ (or $\bar{b}_l$) denotes the $b$ (or
$\bar{b}$) quark (or antiquark) associated with the primary lepton $l^+$
(or $l^-$) from top (or antitop ), while $\bar{b}_j$ (or $b_j$) denotes
the $\bar{b}$ (or $b$) antiquark (or quark) from the decay of the
associated antitop (or top) quark in the top-antitop production and decay
event. The superscript $+$ (or $-$) on these two brackets gives the charge
sign $+$ (or $-$) observed for the primary lepton in this event. 

\smallskip

\noindent \textbf{40758}. This event is of type
$(\bar{b}_l;\bar{q},q,b_j)^+$, since the primary lepton e$^+$ has positive
charge. The SVX tag on jet j1 implies that j1 is either $b$ or $\bar{b}$.

\noindent a) For j1=b, we found no fits, no matter how the other 3 jets
are assigned. 

\noindent b) With j1=$\bar{b}$, the event can only be of type
$(\bar{b}_l;\bar{q},q,b_j)^+$; which jet is $\bar{b}_l$ is to be
determined from fits to the data, considering all 6 identifications for
the other 3 quark jets. We found two fits, one corresponding to the CDF
fit (j4;,j3,j2,j1)$^+$ and having $LIP=3.4$, the other being
(j2;j3,j4,j1)$^+$
but having $LIP=5.4$. 

\smallskip
\noindent \textbf{43096}. This event is of type
$(\bar{b}_l;\bar{q},q,b_j)^-$, the primary lepton being e$^-$. The SVX tag
on jet j1 implies that j1 is either $b$ or $\bar{b}$.

\noindent a) If j1 is b, it can only be $b_j$; we found only one very poor
fit ($LIP=7.3$) and that was for the assignement (j4;j2,j3,j1)$^-$. 

\noindent b) If j1 is $\bar{b}$, it can only be $\bar{b}_l$. Our only fit
was (j1;j3,j4,j2)$^-$, as was found by CDF, here with $LIP=5.5$, larger
by 2 than the most probable $LIP$ values for this batch of events. 

\smallskip
\noindent \textbf{43351}. This event is of type  
$(\bar{b}_l;\bar{q},q,b_j)^-$, the primary muon being $\mu^-$. The only  
fit found was (j2;j1,j3,j4)$^-$. There is a seconday $\mu^-$ from j4,
which suggests that j4 is $b_j$ since the leptonic decay mode for
 b is $b\rightarrow cl^-\nu$; this $\mu^-$ has a p$_T$ relative to j4 of
only 0.46 GeV/c. This is all consistent with the CDF fit. So also is the
fact that $LIP=6.3$, which indicates a poor fit. CDF found $\chi^2$=6.1. 

\smallskip
\noindent \textbf{45610}. This event is of type
$(b_l;\bar{q},q,\bar{b}_j)^+$, the primary muon being $\mu^+$. The SVX tag
on jet j1 implies that j1 is either $b$ or $\bar{b}$. 

\noindent a) If j1 is $b$, it can only be $b_l$. We found only one fit
with this assignment, given by (j1;j3,j4,j2)$^+$ and having $LIP=5.1$,
marginally acceptable. 

\noindent b) If j1 is $\bar{b}$, then j1 can only be $\bar{b}_j$, so we
examined all 6 possibilities $(\bar{b}_l;\bar{q},q,j1)^+$, finding only
one fit, (j2;j3,j4,j1)$^+$. It is very similar to the fit reported by CDF,
and has the very acceptable $LIP$ value of 3.8. 

\smallskip
\noindent \textbf{45705}. This event is of type
$(\bar{b}_l;\bar{q},q,b_j)^-$, the primary lepton being e$^-$. Considering
all possible assignments of the jets, we found two acceptable fits. The
better fit is for (j3;j1,j2,j4)$^-$, with $LIP=3.4$, and it is roughly
comparable with the fit obtained by CDF. This fit is also in accord with
the observed SLT, since this secondary lepton is an e$^+$ with a p$_T$ of
1.61 GeV/c relative to jet j3; the fit requires j3 to be due to a
$\bar{b}$ quark,
for which we have $\bar {b}\rightarrow \bar{c} l^+\nu$. Our second fit was
(j4;j1,j2,j3)$^-$ with a poorer $LIP$ of 5.0, but this reqires j3 to be a
b quark, which disagrees with the SLT, so we reject it.

\smallskip
\noindent \textbf{45879}. This event is of type
$(b_l;\bar{q},q,\bar{b}_j)^+$, since the primary lepton is $\mu^+$. The
SVX tag on jet j2 implies that j2 is either $b$ or $\bar{b}$. 

\noindent a) If j2 is b, then it can only be $b_l$. We found two fits,
(j2;j3,j4,j1)$^+$ with $LIP=3.4$ and (j2;j1,j4,j3)$^+$ with $LIP =6$.
This event also has an SLT, a secondary $\mu^-$ having a p$_T$ of 0.27
GeV/c with repect to j2. 

\noindent b) If j2 is $\bar{b}$, then it can only be $\bar{b}_j$. We found  
one fit in this case, (j1;j3,j4,j2)$^+$ with $LIP=5.3$. This is
essentially  
the same fit as that given by CDF. However, besides gaving a less
acceptable  $LIP$ value, this fit is inconsistent with the SLT $\mu^-$
from  
jet j2, since we know $\bar{b}\rightarrow \bar{c} l^+\nu$. For this CDF
fit to be adopted, it would be necessary to accept this SLT $\mu^-$ as a  
``tertiary lepton''. This would mean that the transition
$\bar{b}\rightarrow\bar{c}$ should be nonleptonic, being then followed by  
the tertiary decay $\bar{c}\rightarrow\bar{s}\mu^-\nu$ which produces the  
observed $\mu^-$ meson. We met a similar, but much more striking, case for  
the dilepton event 41540 in Sec.3, which is discussed
in more detail in our Appendix below. 

\smallskip
\noindent \textbf{45880}. This event is of type
$(\bar{b}_l;\bar{q},q,b_j)^-$, since the primary lepton is e$^-$. All 24
assignments for the 4 jets were examined, and we found two fits,
(j1;j2,j,j3)$^-$ for $LIP=3.4$ and (j2;j1,j4,j3)$^-$ for $LIP=3.5$; the
latter is similar to the fit reported by CDF. Since j3 is the b quark in
both cases, they are both consistent with the observed SLT e$^-$, which
has p$_T$=0.27 GeV/c relative to the jet j3. 

Three of the above events are questionable. Event 43351 has $LIP=6.3$, a
large value corresponding to low probability. It agrees in some detail
with the best
fit given by CDF (Abe, {\textit{et al.}} 1994b) which is however a poor
fit, with $\chi^2=6.1$. This indicates that this event is most probably
not an example of $\bar{t}$-$t$ production and decay and can be rejected
with confidence. Both CDF and our analysis of event 45610 are in accord
on the $m_t$ value, but CDF report $\chi^2=5.0$, which suggests a poor
fit whereas we have $LIP=3.8$ which is indicative of a good fit. On the
other side, for event 43096, where there is good agreement in the $m_t$
value, we have $LIP=5.5$, another large value, whereas CDF find
$\chi^2=2.0$, indicating a good fit. We do not yet understand these
discrepancies.

\section{Conclusions}

    Using QCD at tree level, we constructed by Monte Carlo methods two batches 
of $p\bar{p}\rightarrow t\bar{t}$ production and decay events, for
$p\bar{p}$ center of mass energy 1.8 TeV
and top mass $m_0=170$ GeV, one leading to final states of the type 
$l^{\pm}b\bar{b}\: 2\mathrm{jets}$ with one neutrino (or one
antineutrino),
and the other leading 
to final states of the type ($e^+e^-$ or $e^{\pm}\mu^{\mp}$ 
or $\mu^+\mu^-$) $b\bar{b}$ with two 
corresponding neutrinos, using the simplest tree graph for the first step 
$q\bar{q}\rightarrow t\bar{t}$, followed by the top and antitop decay
sequences. Our methods of 
analysis were then applied to these batches of events, in order to learn what 
outcome we should expect when we apply these methods to real candidate events 
from experiment, which may or may not be correctly interpreted as
$t\bar{t}$ production and decay events.

    The outcome was remarkable. The use of the Bayesian approach led us to a
probability distribution for the mass value $m(i)$ for each event $i$, and
thence to a distribution of the peak mass values $m_{pk}(i)$ for all $i$,
separately for
the two batches. The final distributions for the two batches proved to be 
unexpectedly different, that for the unilepton 
events being sharp and peaking only several GeV below the input mass
$m_0$,
while that for the dilepton-$b\bar{b}$(jets) was very broad and
strongly  asymmetrical. With $m_0=170$ GeV, the latter distribution has
a mean mass of
162.0, its median being about 165.0 GeV. This means that half of the 
dilepton events analysed lead to peak values lying at or below 165 GeV. In
our earlier analyses of real events (Dalitz \& Goldstein 1995), we had  
already noticed this tendency  
for dilepton mass values to lie lower in mass than those for the unilepton
mass values and commented upon it more than once. Now we see that this 
behaviour was to have been expected.

    We made similar calculations for batches constructed for input mass
$m_0=140$~GeV, and some of the corresponding distributions have
been recorded in
the main text. They have some educational value, although being no doubt 
academic.

    Three more detailed comments follow:

\noindent i)   Our analysis of the seven $l^{\pm}4j$ events now known
is in general accord with the CDF-analysis, especially with their
mass estimate of about $175 ~{\rm GeV}$. Two of the events have very low
likelihoods in our analysis, while two of them have relatively
large $\chi^2$ in the CDF analysis, one event being rejected by both;
four events stand firm in both analyses.  The three events rejected
may be due to background such as that originating from the
processes $W^{\pm}+4 jets$ with $W^{\pm} \rightarrow l^{\pm}$,
as discussed by Berends, {\em et al.} 1991, although those
authors show that tagging a single $b(\bar{b})$ quark should significantly
reduce that background. Their calculations indicate a suppression of
background by about
$10^{-2}$ when both $b$ and $\bar{b}$ are tagged.  More estimates from
other mechanisms involving b-quarks need to be considered
quantitatively, within the framework of our analysis procedure.

\noindent (ii)  We have not paid attention to the relative rates for
$l^{\pm}4jets$ and ($e^+e^-$ or $e^{\pm}\mu^{\mp}$ or $\mu^+
\mu^-$)$2jets$
events, and this is an important problem for the future. Accepting that
four b-tagged events
of the former class have been observed, we need to calculate the expected
number of events of the latter class.  This is a complicated calculation,
which is sensitive to the precise cuts which are imposed and which
we do not attempt to carry out here.  The efficiencies depend on
whether the lepton in question is an electron or a muon.  The
nature of the identification given by tagging is different for SVX
and SLT.  SVX does not distinguish $b$ from $\bar{b}$, since it determines
only the location of the secondary vertex, while SLT does not give
the location of the vertex but does distinguish between $b$ and
$\bar{b}$. Since the c-quark decay lifetime is
shorter than that for the b-quark, there should frequently be seen
a tertiary vertex arising from c decay, not far from a secondary b-
vertex.

\noindent iii) Finally, the peak masses $m_t$ determined empirically
appear to be somewhat lower
for $e^{\pm}\mu^{\mp}2jet$ events, on average, than for $l^{\pm}4jet$
events, in accord with the qualitative expectations from our QCD model and
its numerical evaluation. It will be of interest to watch how the
empirical data turn out in future, after the Main Injector comes into
operation at Fermilab. As things now stand, there is no clear discrepency
between our analyses of these two classes of events.

\bigskip
\begin{acknowledgments} 

The authors are grateful to J. Ohnemus for supplying a copy of
an efficient Monte Carlo code with importance sampling and to P. Sphicas,
Tony Smith, K. Kondo, K. Sliwa and D0 group members for many useful
conversations.
G.R.G. thanks the U.S. Department of Energy for partial support of this
research (DE-FG-02-92ER40702), and Prof. John Negele for hospitality at
the MIT Center for Theoretical
Physics during a sabbatical leave when an earlier version of this work was
begun. We appreciate the hospitality of Prof. D. Sherrington at the
Department of Theoretical Physics, Oxford.

\end{acknowledgments}
\bigskip

\begin{appendix}
\section{Appendix. Secondary and tertiary leptons. }

The CDF dilepton event 41540 has a unique interpretation when
analysed as $\mu^+e^-j(1)j(2)$,  $j(1)$ being identified as the $b$-jet and
$j(2)$ as the $\bar{b}$-jet (see Table 1). The jet $j(3)$ is close to the
initial direction, being most probably due to gluon bremsstrahlung. The
event has a third lepton, a ``slow'' $\mu^+$
of energy $8.9 ~{\rm GeV}$. The routine choice of associating this lepton with
the $\bar{b}$ jet is
not convincing since its largest momentum component, $p_x(\mu^{+}) = 8.7
{}~{\rm GeV}$, is oppositely directed to the largest component of the jet 2
momentum, $p_x = -50.0 ~{\rm GeV}$.
It is much more plausible that the ``slow'' $\mu^+$ is associated with jet 1,
since its momentum is almost parallel with the momentum of jet 1, and in
the same direction; its momentum $p_\perp$, transverse to jet 1 is only
about $0.6 ~{\rm GeV/c}$.

The sequence of quark processes which lead to the emission of a tertiary
$\mu^+$ lepton, (A2) with, and (A3) without, a secondary $\mu^-$ lepton,
are
as follows:
\begin{equation} t \rightarrow b + W^{+} \hspace{2in} W^{+}\rightarrow
\mu^{+}+\nu_{\mu}, \label{3a}
\end{equation}
\begin{equation} \hspace{0.5in} b \rightarrow c +W^{-} \hspace{1.2in}
W^{-} \rightarrow \mu^{-}+\bar{\nu}_{\mu} \label{3b}
\end{equation}
\begin{equation}
\hspace{1.5in} {\mathrm{or}}
\hspace{0.5in}
W^{-} \rightarrow {\mathrm{hadrons}}(\bar{u} d + \bar{c} s),
\label{3c}
\end{equation}
\begin{equation} \hspace{1.0in} c \rightarrow s + W^{+} \hspace{1.0in} W^{+}
\rightarrow \mu^{+} + \nu_{\mu}, \label{3d}
\end{equation}
the W's in (\ref{3b}), (\ref{3c}) and (\ref{3d}) being necessarily virtual,
of course.
The net process for the event 41540 would be the consequence of
(\ref{3a}),(\ref{3c}) and (\ref{3d}), thus:
\begin{equation} t \rightarrow {\mathrm{hadrons\/}} +
\mu^{+}({\mathrm{fast}}) + \mu^{+}({\mathrm{slow}}) + {\mathrm{neutrinos}}.
\label{4}
\end{equation}

To orient ourselves concerning the final states, we have made some simple
model calculations for the momentum distributions for a secondary lepton
$l_2$, or for  a tertiary lepton $l_3$, appropriate for an initial
$b$-jet
with momentum about $130 ~{\rm GeV/c}$. We adopted the fragmentation
functions
of Peterson {\em et al.} (1983), with the parameter
$\epsilon_Q=(0.49)/m_Q^2  ~{\rm GeV}^{-2}$, where $m_Q$ denotes the appropriate
heavy quark mass. In the first step, the $b$-quark generally undergoes
hadronization to a final ground state meson $B_u^{-}$, $B_d^{0}$ or
$B_s^{0}$ with spin-parity $0^{-}$, together with some number of light
mesons; we neglect explicit mention of hadronization to $\Lambda_b$
baryons, since such final states contribute much less and do not affect
the over-all conclusions. Using the standard model expression for the
momentum distribution of the lepton resulting from $b\rightarrow c l_2
\nu_l$ in the B-meson rest frame, we obtain the $l_2^{-}$ momentum
distribution in the lab frame by integrating this distribution over the
B-momentum distribution given by the fragmentation function. The
resulting energy distribution for secondary leptons is given in
Fig.10(a).
We note that these energies run up to very large values. The mean
$l_2^{-}$ energy is $\approx 33 ~{\rm GeV/c}$ and $50\%$ of the leptons have
energy greater than $29 ~{\rm GeV}$. The distribution for $p_{\perp}$, 
the secondary
muon momentum transverse to the B-momentum, is given in Fig.11(a),
although we must note that the B-momentum is affected by the gluons and
light mesons emitted so that it differs a little from the $b$-jet axis
observed. The most probable value for $p_{\perp}(l_2)$ is 
$1.4 ~{\rm GeV/c}$; its
median value is $1.35 ~{\rm GeV/c}$. For $90\%$ of the secondary leptons,
$p_{\perp}(l_2)$ exceeds $0.6 ~{\rm GeV/c}$; 
$70\%$ of them have $p_{\perp}(l_2) \geq 1 ~{\rm GeV/c}$.
For tertiary leptons, we must first carry out the same calculation for
the lab momentum distribution of the c-quarks from the b-jet. Naturally,
this distribution is quite different from that for the secondary leptons,
because of the large mass value for the c-quark. The lab energy
distribution for the $l_3^{+}$ lepton from $c\rightarrow s l_3^{+}\nu_l$
decay is then obtained by integrating the latter, as given by the
standard model, over the fragmentation function for the ground state
$0^{-}$ ($D_u, D_d \:{\mathrm{and}}\: D_s$)-mesons from the c-jet
distribution
just calculated. The resulting energy distribution for the $l_3^{+}$
lepton is shown in Fig.10(b). We note that these energies are much less
than those for secondary leptons but their distribution is very
asymmetric; their peak value is $\approx 0.5 ~{\rm GeV}$, while their median
value is $\approx 5 ~{\rm GeV}$. Above $1 ~{\rm GeV}$, 
the distribution falls gradually
with increasing energy $E(l_3)$, by a factor of 3 from 5 to 15 GeV, and
then faster beyond; $25\%$ of the tertiary leptons have $E(l_3)\geq 10
{}~{\rm GeV}$, but only $9\%$ have energies exceeding $15 ~{\rm GeV}$. 
The distribution for
the transverse momentum $p_{\perp}(l_3)$ is shown on Fig.11(b). It peaks
at $0.35 ~{\rm GeV/c}$ and is a little asymmetric; about $30\%$ of the
events have
$p_{\perp}(l_3) \geq 0.6 ~{\rm GeV/c}$, 
about $5\%$ have $p_{\perp}(l_3) \geq 1.0 {}~{\rm GeV/c}$.

\begin{figure}
\begin{center}\leavevmode\epsffile{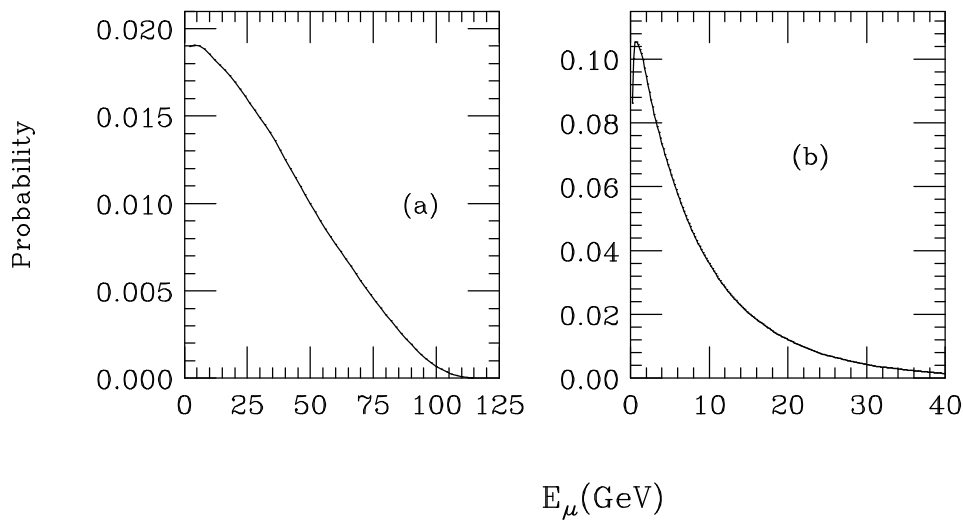}\end{center}
\caption{(a) The energy distribution in the Lab. frame, for the secondary
leptons resulting from the decay $b\rightarrow c+l^{-}+\nu_l$, for a b-quark
jet of initial energy $130 ~{\rm GeV}$; (b) The energy distribution in
the Lab. frame, for tertiary
leptons resulting from the decay $c\rightarrow s+l^{+}+\nu_l$, for a b-quark
jet of initial energy $130 ~{\rm GeV}$.}
\end{figure}

\begin{figure}
\begin{center}\leavevmode\epsffile{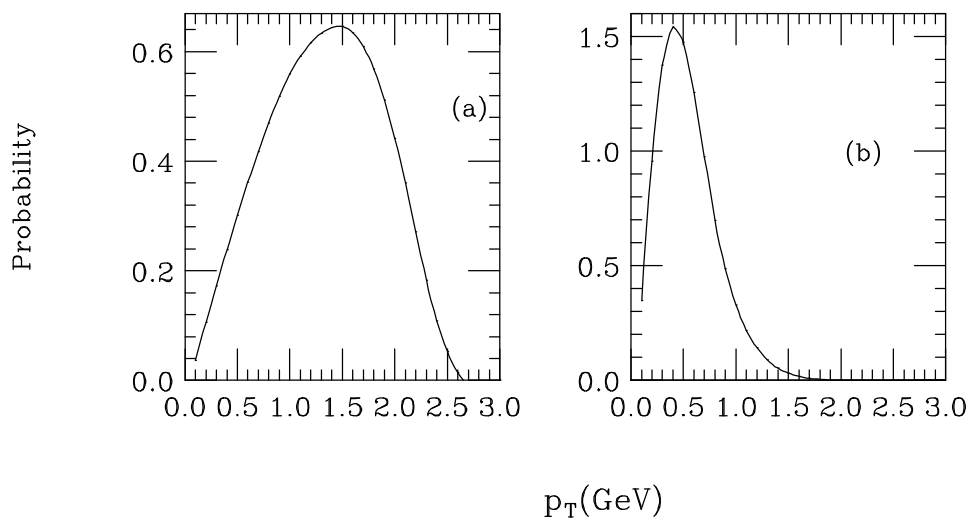}\end{center}
\caption{(a) The distribution of the momentum transverse to the b-jet axis,
for secondary leptons resulting from the decay $b\rightarrow
c+l^{-}+\nu_l$, for a b-quark jet of initial energy $130 ~{\rm GeV}$;
(b) The distribution of the momentum transverse to the b-jet axis,
for tertiary leptons resulting from the decay $c\rightarrow
s+l^{+}+\nu_l$, for a b-quark jet of initial energy $130 ~{\rm GeV}$. }
\end{figure}

We now return to the consideration of event 41540. That the
``slow'' $\mu^{+}$ lepton is associated with jet 1
is supported by a close examination of the event shown in Fig.10 and
Table VII of the CDF paper (Abe {\em et al.} 1994b). There is a displaced 
vertex shown in the SVX detector, at
$\bar{r}=0.33 ~cm$ from the origin of the event.
Comparison of the $\phi$ distribution in their Fig.10(b) with the entry
in their
Table VII shows us that the secondary vertex shown is associated with the b-jet
(jet 1). We are not told where the $8.9 ~{\rm GeV/c}$ $\mu^+$ emerged.
The two most immediate possibilities are:

(T$_1$) the displaced vertex is a non-leptonic b-decay, the ratio
$\bar{r}/\bar{d}_B$ being 0.26, where $\bar{d}_B = \gamma_B c \tau_B$
is the mean distance of travel by the b-quark before decay,
$\gamma_B m_B = 131 ~{\rm GeV/c}$, and $\tau_B$ being the B-meson lifetime.
The resulting c-quark then undergoes decay $c \rightarrow s \mu^{+}
\nu_{\mu}$, leading to a ``slow'' $\mu^+$ which is tertiary. The chance
that this c-decay occurs outside the SVX region is about
$e^{-((d-\bar{d}_B)/\bar{d}_D)}$
where $d=0.5 ~cm$ and $\bar{d}_D =\gamma_D c \tau_D$. Taking $\gamma_D$ to
have value about the same as $\gamma_B$, we then have $\bar{d}_D$ about
$0.37 ~cm$, which gives the chance of the c-quark escaping without detection
to be about $40\%$.

(T$_2$) The vertex observed is a tertiary decay, the ``slow''
$\mu^+$ being
one of the tracks observed (whether or not it is identified) and coming from
the transition $c \rightarrow s\mu^{+}\nu_{\mu}$. The only question is
``where is the b-quark decay vertex?'' To give rise
to what is observed, there should then be a b non-leptonic vertex between the
origin and the displaced vertex, but perhaps so close to the tertiary
vertex, in view of the rapidity of c-decay relative to b-decay, that it may
be difficult to separate the two vertices. Also, in this case, there should
necessarily be a ``slow'' $\mu^+$ emitted from the displaced vertex, although  
there is no clear statement identifying this    
$\mu^+$ in the SVX data. It is
difficult to estimate the probability for this outcome, without more detailed
information. A much closer examination of the SVX data on this event is needed.

Such tertiary leptons will not be rare.
The branching fraction (BF) for all leptonic modes is known
(Particle Data Group 1996) to be about $21.0(4)\%$ for the b quark and
about $23(3)\%$, on
average, for the c quark, assuming that the configurations
$(\bar{u} c), (\bar{d} c) \,{\mathrm{and\/}}\, (\bar{s} c)$ are produced
equally often. (For the $D$ mesons, the BF's are
$34.4(38)\%$ for $D^+$ and $17.7(24)\%$ for $D^0$. From their known
lifetimes their leptonic decay rates are therefore
$3.3(4)\times~10^{11} ~s^{-1}$ and $4.3(7)\times~10^{11} ~s^{-1}$,
respectively, in fair agreement with each other. The well known inequality
between their total decay rates (and therefore between their leptonic
BF's) is due to a suppression of the non-leptonic decay modes of $D^+$
relative to those for the $D^0$. However, it is leptonic BF's which are
relevant for discussing the possibilities for tertiary leptons. The leptonic
BF's are not known for $D_s^+$, only the upper limit, $<20\%$, but its
total lifetime is within three standard deviations of that for $D^0$.)
Neglecting corrections for the efficiencies for detecting SLT's, generally
stated to be about $30\%$ but which may be substantially lower than
this for the detection of tertiary leptons, we may estimate that
the frequency of tertiary leptons without any secondary lepton is
comparable with the frequency of secondary leptons without any
tertiary lepton.

However, there is an alternative interpretation possible for the ``slow''
$\mu^{+}$:

(S$_d$ or S$_s$) The hadronization of the b-jet may lead to a charged
$B^{-}$ meson or to a neutral meson, $B_d^0$ or $B_s^0$. In the latter
two cases, the meson may
undergo the process of $(B_d^0,\bar{B}_d^0)$ mixing or $(B_s^0,\bar{B}_s^0)$
mixing,
and can then emit a $\mu^+$ lepton from the secondary process
$\bar{b} \rightarrow \bar{c} \mu^+ \nu_{\mu}$, since this will be possible
from the
$\bar{b}$-quark in the $\bar{B}_d^{0}$ or $\bar{B}_s^{0}$ components of
the final mixed ($\bar{B}^{0},B^{0}$)-meson state.
{}From data on b-jet development following the much studied process
$Z^0 \rightarrow b\bar{b}$, it is
known that the secondary $\mu^+$'s from this source have an intensity
$13\%$ of the total from the secondary ($\mu^{+}+\mu^{-}$) leptons from
the  initial
b-quark. These secondary $\mu^+$'s from mixing will have the same energy
spectrum as the $\mu^{-}$ secondary leptons from all three kinds of final
B-meson, which we have estimated from our model calculation to have the
form shown in Fig.10(a), a spectrum much harder than our estimate for the
tertiary $\mu^{+}$ spectrum, given in Fig.10(b).

We may now our calculated probability curves to assess the
relative likelihood of the two hypotheses, T and S, just discussed above.

(S)  $b\rightarrow \bar{b}\rightarrow \bar{c} l^{+}$ and $b\rightarrow c
l^{-}$.

As noted above, it is known (Particle Data Group 1996) that the rate for
$l^{+}$ is
$\epsilon = 0.13$ times that for ($l^{+}+l^{-}$) when the sum is over
$B_d^{0}$ and $B_s^{0}$ mesons. We denote the distribution of the final
secondary lepton by $P_2(E_l)$, shown in Fig.10(a), and the distribution
of the secondary lepton momentum transverse to the b-jet axis by
$Q_2(p_{l\perp})$, shown in Fig.11(a). From the Particle Data Group 1996, 
we take
$B_{bl}=0.207$ for the branching fraction $(b\rightarrow {\mathrm{all}} \:
l^{\pm})/({\mathrm{all}}\ b\ {\mathrm{decays}})$. The net rate for
$l^{+}$, occuring as secondary leptons, is given by
\begin{equation}
R_S = \epsilon\cdot B_{bl}\cdot P_2(E_l)\cdot Q_2(p_{l\perp}).
\label{5}
\end{equation}
per inital b quark.

(T)  $b\rightarrow c \rightarrow l^{+}$, with no secondary lepton.

Here we ignore the SVX detector, i.e. we do not require the second decay
to be visible within it. We take $B_{cl}=0.34$ (Particle Data Group 1996)
as the branching fraction ($c\rightarrow {\mathrm{all}}\:
l^{\pm})/({\mathrm{all}}\: c\:
{\mathrm{decays}})$. The net rate for $l^{+}$ is now,
\begin{equation}
R_T = (1-B_{bl})\cdot B_{cl}\cdot P_3(E_l)\cdot Q_3(p_{l\perp}).
\label{6}
\end{equation}
per initial b quark.

For the event of interest, we have $E_{\mu}=8.9 ~{\rm GeV}$ and
$p_{\mu\perp}=0.60 ~{\rm GeV/c}$. 
The interpretation S that the ``slow'' $\mu^{+}$
lepton is due to ($\bar{B}^{0},B^{0}$) mixing gives the rate per initial b
quark as
\begin{equation}
R_S=0.22\times 0.13\times 0.018\times 0.36 = 4.4\times 10^{-4}.
\label{7}
\end{equation}
With the tertiary interpretation T, we have the rate
\begin{equation}
R_T=0.78\times 0.33\times 0.035\times 1.24 = 1.12\times 10^{-2}.
\label{8}
\end{equation}

Hence the calculations with our simple model for the decay sequences
$b\rightarrow c l^{+}\nu$ and $b\rightarrow c\rightarrow s l^{+} \nu$
indicate that the likelihood that this $\mu^{+}$ is tertiary relative to
the likelihood that it is secondary - but results from ($\bar{B}^{0},B^{0}$)
mixing - is 25:1. The main factor depressing the rate $R_S$ is the low value
for $\epsilon$; surprisingly, the observed values for $E_l$ and
$p_{l\perp}$ do not distinguish clearly between the possibilities S and T.

It is of interest to compare event 41540 with those SLT
$l4\mathrm{jets}$ events reported by CDF, which can be assigned uniquely
and kinematically to secondary lepton emission. These are the events where
the lepton charge has sign in accord with the decay $b\rightarrow c l^{-}
\bar{\nu}$ (or $\bar{b}\rightarrow \bar{c} l^{+} \nu$, for events which
stem from $\bar{t}$ production and decay). We note that the primary lepton
energies $E_{1l}$ have a reasonable spread of energies, from $24.7$ to $117.3
{}~{\rm GeV}$, as shown in Table 3. 
The two energies $E_{1l}$ in the dilepton
events lie at energies in the range $\approx 30-70 ~{\rm GeV}$. The four  
``slow'' leptons available have energies $E_{2l}$ which range from $2.4$
to $14.3 ~{\rm GeV}$,
while the ``slow'' $\mu^{+}$ energy in event 41540 lies in the middle of
this range. The same holds for its $p_{l\perp}$ value. Since the b-jet
energy in this event has a surprisingly large value, ($E_{bj}+E(SLT)$)
being $\approx 141 ~{\rm GeV}$ (overlooking the unknown neutrino energy
resulting from this b-decay), we might look instead at the weighted
energies $E(SLT)/(E(SLT)+E_{bj})$ and transverse momenta
$p_{l\perp}/(E(SLT)+E_{bj})$, listed in Table 5. Even then, their values
for 41540 still lie within the ranges obtained for these parameters
from the four $l4\mathrm{jets}$ events. None of these numbers mark out this
SLT event as being obviously different from the other SLT events, except
for the charge sign for the ``slow'' $\mu^{+}$ and the magnitude of the
ratio $R_T/R_S$ discussed above.

\begin{table}
\caption{Events available having a secondary or tertiary lepton} 
\longcaption{Energies
in GeV, momenta in GeV/c. Bracket denotes possible tertiary lepton.}
\begin{tabular}{|c|r|r|r|r|r|} \hline
Event                        & 43351 &45705      & 45879     &
   45880          &    41540                     \\ \hline
$E_{1l}$                     &  24.7 &   70.0    & 54.0      &
   26.1           &   48.2                         \\ \hline
$E_{bj}$                     &  39.0 &   70.7    & 96.7      &
   46.5           &   131.8                      \\ \hline
$E_{2l}$                     &  2.37 &  14.28    & 13.58     &
   2.80           &   (8.9)                      \\ \hline
$E_{2l}/(E_{2l}+E_{bj})$ & 5.7\% &     16.0\%    & 12.6\%    &
   5.7\%          &   (6.8\%)                    \\ \hline
$p_{l\perp}$             &  0.46 &     1.615     &  0.27     &
   0.27           &    0.60                      \\ \hline
$p_{l\perp}/(E_{2l}+E_{bj})$ & 1.1\% & 1.45\%    & 0.25\%    &
   0.55\%         &   (0.43\%)                   \\ \hline
\end{tabular}
\end{table}

Finally, we must compare these four SLT $l4\mathrm{jets}$ events with the
calculated spectra for our simple
($b\rightarrow c l^{-}\bar{\nu},\: \bar{b}\rightarrow \bar{c} l^{+} \nu$)
model. Fig. 10(a) shows that the median value
predicted for secondary lepton energy $E_{2l}$ lies at 
$\approx 30 ~{\rm GeV}$ for
the dilepton event 41540, with $130 ~{\rm GeV}$ for jet 1 lab energy
$E_{bj}$. The four SLT events have lower b-jet energies, ranging
from $39$ to $97 ~{\rm GeV}$. This does not
effect the $p_{l\perp}$ spectrum, but it alters the $E_{2l}$ spectra. For
each event the energy spectrum
will depend on the boost from the decaying B rest frame to the lab frame
(in which the B meson is a fragment of the b-jet).
For the event with the lowest associated b-jet energy, event 43351, the
corresponding spectrum will have a median of about $9 ~{\rm GeV}$, compared to
the measured $E_{2l} = 2.37 ~{\rm GeV}$. 
The median energy grows roughly linearly with jet energy, so
the secondary lepton in each of these four events have generally
lower energy, $E_{2l}$, than the predicted median.
Three of these SLT events have
$p_{l\perp}$ values that lie below $0.5 ~{\rm GeV/c}$, whereas our 
model predicts its
median to be $\approx 1.35 ~{\rm GeV/c}$. The accord of these data with our
calculations is neither striking nor unfavourable. It may be that the cuts made
on the data by the experimenters, aimed at picking out any background
events, have a much larger effect on the predicted curves in Figs.10 and
11 than we have anticipated.

Double-tagging, the combination of the secondary vertex detector (SVX)
and the observation of secondary leptons (SLT) together should provide
a powerful means
for interpretation of the nature of individual events, without a full
dynamical analysis (which would at best be possible only rarely).
The above analysis of event 41540 illustrates this point quite
strongly.
\end{appendix}

\label{lastpage}
\received{Received }
\end{document}